\documentclass[]{jfm}
\usepackage{float}
\usepackage{caption}
\usepackage{graphicx}
\usepackage{newtxtext}
\usepackage{newtxmath}
\usepackage{natbib}
\usepackage{hyperref}
\hypersetup{
    colorlinks = true,
    urlcolor   = blue,
    citecolor  = black,
}

\newcommand{\RomanNumeralCaps}[1]
\linenumbers

\title{Hydrodynamic modulation via cupping in a crustacean-inspired propulsor}
\author[Oliveira Santos et al.]{
S. Oliveira Santos\aff{1,2},
M. Brown\aff{1,3},
M. Kim\aff{1,4},
N. Tack\aff{1},
\and \ M.M. Wilhelmus\aff{1}\corresp{\email{mmwilhelmus@brown.edu}}}

\affiliation{
\aff{1}Center for Fluid Mechanics, School of Engineering, Brown University, Providence, RI 02912, USA
\aff{2}Institute for Geophysics, University of Texas at Austin, Austin, TX 78758, USA
\aff{3}Graduate Aerospace Laboratories, California Institute of Technology, Pasadena, CA 91125, USA
\aff{4}School of Mechanical Engineering, Kookmin University, Seoul 02707, Republic of Korea}

\begin{document}
\maketitle

\begin{abstract}
Shrimp, like many invertebrates swimming at intermediate Reynolds numbers ($Re$), rely on the interplay between morphology and kinematics to generate thrust while producing sufficient lift to overcome their negative buoyancy. Shrimp pleopods branch into an endopodite and an exopodite, whose relative motion varies the projected surface area during the swimming cycle. For this mechanism to function, the exopodite must be cambered relative to the endopodite at a set cupping angle $\zeta$, which partially decouples the effective angle of attack of the exopodite from the overall leg kinematics. Here, we investigate the role of $\zeta$ in modulating thrust--lift balance during steady forward locomotion. Using a dynamically scaled (40$\times$) robotic pleopod, we systematically varied $\zeta$ from $0^\circ$ to $80^\circ$, measured hydrodynamic forces, and performed particle image velocimetry at $Re = 968$. Moderate cupping angles ($\zeta = 20^\circ$--$40^\circ$), consistent with biological observations, provide optimal thrust--lift balance. At these angles, the exopodite abducts rapidly during the power stroke, maximizing projected area at peak flow velocity, and adducts early during the return stroke, minimizing resistive drag. A reduced-order force model reveals that the exopodite contributes 52--62\% of total lift, particularly at intermediate $\zeta$, where a leading-edge vortex (LEV) forms and remains attached throughout the power stroke. At extreme cupping angles, LEV coherence degrades and force production weakens. These findings demonstrate that shrimp pleopods function as hybrid propulsors exploiting both drag- and lift-based forces, and that $\zeta$ serves as a geometric control parameter capable of tuning thrust--lift balance independently of stroke kinematics.

\end{abstract}

\begin{keywords}
Authors should not enter keywords on the manuscript, as these must be chosen by the author during the online submission process and will then be added during the typesetting process (see \href{https://www.cambridge.org/core/journals/journal-of-fluid-mechanics/information/list-of-keywords}{Keyword PDF} for the full list).  Other classifications will be added at the same time.
\end{keywords}

\section{Introduction}
\label{sec:headings}

%Many of the most abundant aquatic invertebrates swimming in the intermediate Reynolds number (Re) regime employ metachronal propulsion, characterized by the sequential beating of several closely spaced appendages whose phase lag induces a continuous traveling wave \citep{morgan1972swimming}. This locomotor mode is generally regarded as a drag-based strategy that relies on the propulsors pushing against water to generate thrust. Spatiotemporal asymmetries, such as bending, morphing, and modulating the duration of the power and return strokes, generate net thrust necessary for propulsion \citep{herrera2021spatiotemporal, tack2025going, murphy2011metachronal, connor2025hydrodynamics, lou2024hydrodynamics}. By impulsively starting their propulsors, metachronal swimmers leverage these mechanisms to produce large force peaks over short durations, making this swimming mode well-suited for acceleration and maneuvering \citep{vogel2013comparative, kim_characteristics_2011}. 

Metachronal locomotion is a swimming strategy characterized by the sequential beating of closely spaced appendages whose phase lag produces a continuous traveling wave \citep{murphy2011metachronal}. Employed by a wide range of aquatic invertebrates — from ctenophores and copepods to shrimp and krill — this locomotor mode operates in the intermediate Reynolds number (Re) regime and is classified as drag-based. The propulsors modify their surface area to minimize drag and generate thrust in the swimming direction \citep{kim_characteristics_2011, herrera2021spatiotemporal, tack2025going}. The production of large force peaks over short periods allows swimmers to accelerate quickly and achieve high maneuverability \citep{vogel2013comparative, kim_characteristics_2011}. These attributes have motivated growing interest in metachronal propulsion as a paradigm for expanding the capabilities of autonomous underwater vehicles (AUVs), particularly for low-speed maneuvering and operation in sensitive or confined environments where conventional propellers risk entanglement or sediment disturbance. However, translating biological performance into engineered systems requires a mechanistic understanding of how metachronal swimmers modulate hydrodynamic forces. While net thrust production through spatiotemporal asymmetries is relatively well characterized \citep{herrera2021spatiotemporal, tack2025going, murphy2011metachronal, connor2025hydrodynamics, lou2024hydrodynamics}, the mechanisms by which these organisms balance thrust with lift remain poorly understood.

%Negatively buoyant metachronal organisms like shrimp must produce thrust during swimming but also generate sufficient lift to overcome their weight \citep{Kils1981}. This requirement has led to specialized morphological adaptations of the swimming appendages (pleopods), which branch from a rigid stalk (protopodite) into two distinct flexible ramal structures, the endopodite and the exopodite (figure \ref{fig:shrimp_morphology}). During the power stroke, both rami remain stiff. The endopodite moves along the vertical plane, similar to a flat plate, forming a tip vortex on its anterior face \citep{kim_characteristics_2011}. In contrast, the exopodite abducts into the free stream by rotating laterally about the ramal joint, causing the exopodite to face the flow with an angle of attack ($\delta$) that promotes the formation of a leading-edge vortex (LEV) and trailing-edge vortex (TEV) \citep{murphy2013hydrodynamics}. These structures suggest that shrimp may exploit LEV stabilization strategies similar to those seen in flying insects, where rotational motion and spanwise flow support vortex attachment \citep{lentink2009rotational}.

\begin{figure}
    \centering
    \includegraphics[width=0.5\linewidth]{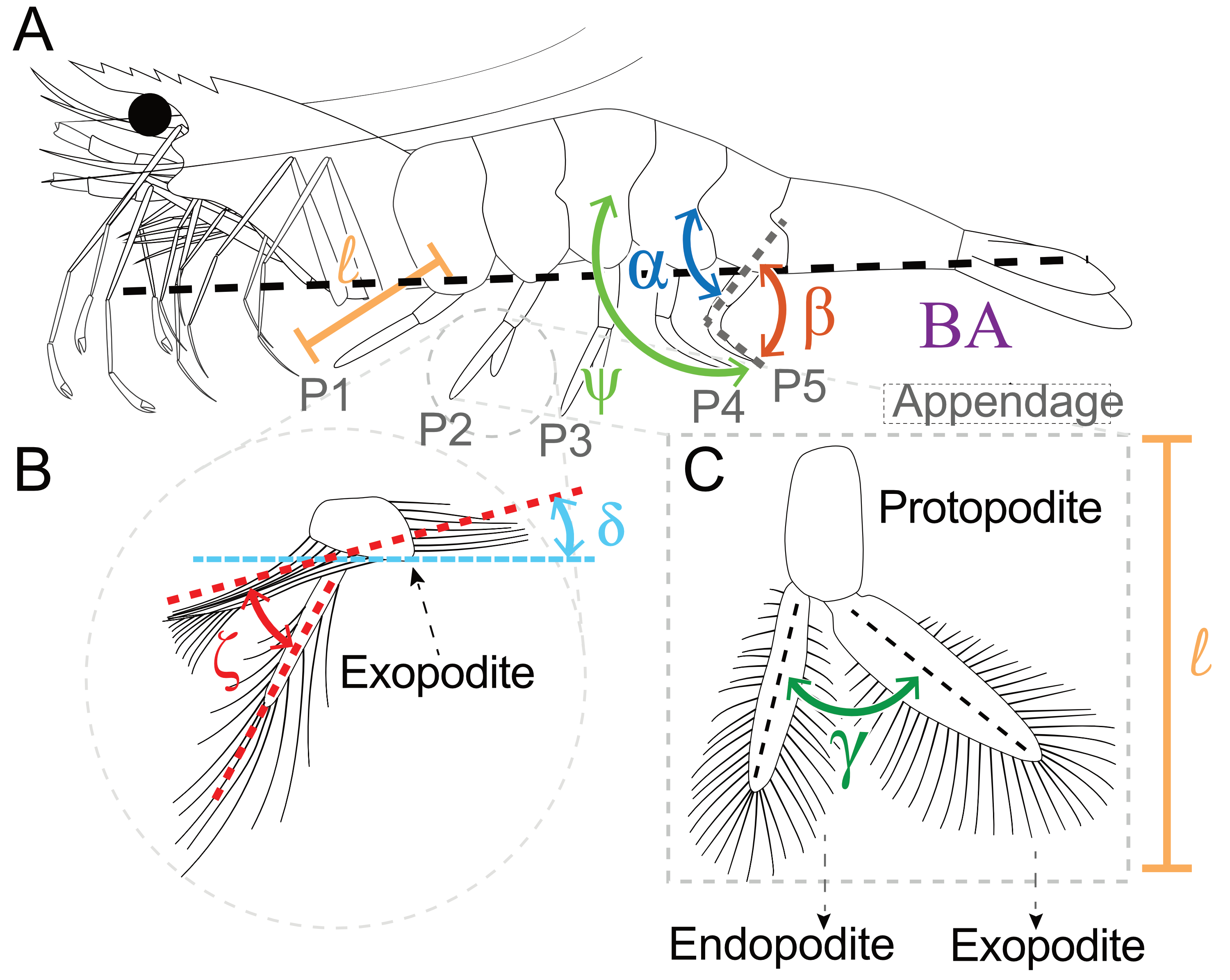}
    \caption{
    \raggedright
    \textbf{Geometric parameters governing shrimp kinematics and morphology.} Labeled parameters are seen in three distinct views. (A) The lateral view shows appendage length, $l$, the angle $\alpha$ between the protopodite (proximal part of pleopod) and the body axis (BA) passing through the roots of the five pleopods, the leeward angle $\beta$ between the protopodite and endopodite, and the $\psi$ angle between BA and the endopodite. (B) The zoomed lateral view of a pleopod shows the cupping angle, $\zeta$, which is the camber angle of the abducted exopodite relative to the endopodite. (C) The frontal view of a pleopod shows the out-of-plane abduction angle, $\gamma$.}
    \label{fig:shrimp_morphology}
\end{figure}

Many aquatic organisms are negatively buoyant. To regulate their vertical position in the water column, they must generate lift while keeping the energetic costs low. Teleost fish, for example, have gas-filled swim bladders \citep{jones1953structure}. In contrast, shrimp and krill produce lift in part through appendage kinematics, placing additional demands on the swimming appendages (pleopods) beyond thrust generation alone. This dual requirement appears to have driven the evolution of specialized pleopod morphology as each pleopod branches from a rigid stalk (protopodite) into two distinct structures, the endopodite and the exopodite (figure \ref{fig:shrimp_morphology}). During the power stroke, both rami remain stiff. The endopodite sweeps fluid about the body line, functioning similarly to a flat plate and forming a tip vortex on its anterior face \citep{murphy2013hydrodynamics, connor2025hydrodynamics}. Conversely, the exopodite abducts into the free stream by rotating laterally about the ramal joint. As a result, the exopodite faces the oncoming flow at an angle of attack ($\delta$), promoting the formation of a leading-edge vortex (LEV) and a trailing-edge vortex (TEV) \citep{murphy2013hydrodynamics, oliveira2023pleobot}. 

The presence of these vortical structures suggests that shrimp exploit lift-enhancing mechanisms analogous to those observed in insect flight. In flapping wings, for example, formation and attachment of a leading-edge vortex increase circulation about the lifting surface, establishing a sustained low-pressure region on the suction side that augments lift relative to steady-state flow conditions. Rotational motion and spanwise flow contribute to vortex stabilisation, delaying detachment and maintaining elevated circulation throughout the stroke \citep{lentink2009rotational}. By abducting into the free stream, the exopodite generates comparable unsteady flow conditions in which the LEV increases circulation and produces a low-pressure region that contributes to lift generation. The two rami of shrimp pleopods therefore support thrust production while regulating vortex dynamics to sustain lift with limited energetic cost.

Shrimp modulate thrust and lift through several mechanisms, including varying the stroke amplitude of their pleopods and leveraging the cupping angle ($\zeta$) between their appendages \citep{murphy2013hydrodynamics, martin2018role, connor2025hydrodynamics, herrera2021spatiotemporal}. The cupping angle is the dihedral angle between the exopodite and endopodite surfaces. It sets the initial incidence of the exopodite relative to the endopodite plane. During the power stroke, appendage rotation combined with exopodite pitching produces an instantaneous $\delta$ between the exopodite chordline and its local direction of motion. This angle of attack varies continuously throughout the stroke, and is set through the combined effects of pleopod rotation, abduction, and cupping. Hence, $\zeta$ provides a mechanism to partially decouple the effective angle of attack from the overall leg kinematics. This enables shrimp to optimize force production without substantially altering stroke timing or amplitude.

Simplified models such as translating and flapping plates have served as analogues to analyse the role of distinct vortex structures during drag-based propulsion \citep{kim_characteristics_2011, fernando2016reynolds, fernando2016vortex, chen2020leading}. At intermediate Reynolds numbers, boundary layers are prone to separation but can be stabilized by rotational and spanwise flows \citep{eldredge2019leading}. LEVs may thus play a central role at these Reynolds numbers. Not only can they stabilize the boundary layer, but they can also promote lift \citep{venkata2013leading, baik2012unsteady}. The onset and persistence of LEVs depend on several factors, notably propulsor flexibility and aspect ratio \citep{kim_characteristics_2011, fernando2016vortex}. Appendages with  aspect ratios AR $>$ 1 promote early-stage vortex growth and generate the large, transient force peaks characteristic of drag-based propulsion \citep{fernando2016reynolds}. For low-aspect-ratio foils, spanwise flows stabilize LEVs and enhance lift throughout the stroke cycle under flapping or oscillating motions \citep{baik2012unsteady, anderson1998oscillating, wang2015lift}. 

Similar unsteady mechanisms during insect flight (including delayed stall, rotational lift, and wake capture) yield lift-to-drag (L/D) ratios that far exceed steady-state predictions \citep{sane2003aerodynamics, wang2015lift}. In particular, \citet{lentink2009rotational} demonstrated that rotational accelerations can prevent vortex shedding and maintain attached LEVs on revolving insect wings. These mechanisms are not unique to flying organisms: flat plates undergoing spanwise oscillations have demonstrated L/D ratios approaching 1.4 at moderate angles of attack (15–25$^{\circ}$), comparable to values measured in animals \citep{wang2015lift}. The higher-aspect-ratio exopodite of shrimp pleopods, which abducts laterally into the flow at angles of attack similar to those of biological flyers, may exploit similar mechanisms to generate attached LEVs augmenting both thrust and lift.

Formation of an attached LEV stabilizes the flow and enhances force production, likely enabling shrimp to modulate thrust and lift during swimming. The role the cupping angle plays in achieving a $\delta$ that promotes the formation of an LEV and its effect on the resulting forces has not been quantified. Because of the combination of a fixed endopodite and abducting exopodite, shrimp appear to operate metachronal propulsion in a hybrid regime exploiting both drag- and lift-based mechanisms to optimize swimming, whereby thrust and lift result from the pleopods sweeping through the water and exploiting LEVs. Drag-based swimming can still generate vortices, but these structures are typically associated with resistive separation rather than attached, force-enhancing LEVs. 

In this study, we focus on the role of the cupping angle by using a dynamically scaled robotic pleopod model based on the Pleobot architecture \citep{oliveira2023pleobot}, which integrates variable cupping angle and a six-axis force transducer for whole-appendage force measurement. By systematically varying $\zeta$ (see figure \ref{fig:shrimp_morphology}) and measuring time-resolved forces alongside two-dimensional flow fields obtained via particle image velocimetry (PIV), we isolate the contributions of the exopodite to force production and vortex dynamics. We complement the experimental results with a reduced-order force model based on Morison’s equation, extended to account for pleopod rotation and time-varying projected area. Through this combined approach, we show that moderate cupping angles produce a balance of lift and thrust comparable to strategies used in flying insects. Importantly, we demonstrate that the exopodite is critical to lift production by promoting the formation and persistence of attached LEVs. Our findings provide new insights into the fluid–structure interactions underlying metachronal swimming in crustaceans and offer design principles for bio-inspired underwater propulsion systems.

\section{Experimental setup}
\subsection{Robotic analog}
\emph{P. vulgaris} morphometric data were used to design a dynamically scaled (40$\times$) robotic shrimp pleopod based on the \textit{Pleobot} architecture \citep{oliveira2023pleobot}. The pleopod reproduced the membranous portions of the endopodite and exopodite of \emph{P. vulgaris} while excluding the marginal setae (fine hair-like structures). The pleopod analog was fabricated using stereolithography (SLA)  (Saturn 2, Elegoo) and programmed to replicate pleopod kinematics measured in free-swimming shrimp \citep{tack2025going}. Leg motions from a single appendage (P4) were selected as the target kinematics to control the oscillations of the $\psi$ angle (see figure \ref{fig:shrimp_morphology}) with a single servomotor (positioning error = 0.15$^{\circ}$) mounted above the waterline. Note that although the robotic design and primary kinematics are based on \textit{P. vulgaris}, validation of the passive exopodite abduction dynamics $\gamma$ was performed using kinematic data from \emph{P. paludosus}, for which time-resolved $\gamma$ measurements were available. The appendage Reynolds number is defined as
\begin{align}
    Re = \frac{U_{\text{tip}} W_{\text{tot}}}{\nu}= \frac{ (2\pi f\psi_m \overline{l}) W_\text{tot}}{\nu_w},
\end{align}
where $W_{\text{tot}}$ is the total width of the appendage ($W_{exo} + W_{endo}$), $U_{\text{tip}}$ is the tip speed of the appendage, and $\nu$ is the kinematic viscosity of the working fluid. The tip velocity was calculated using the pleopod beat frequency, $f$, the stroke amplitude, $\psi_m$ (in radians), and the average shrimp appendage length, $\overline{l}$. All robotic experiments were performed in a glycerin-water mixture (60$\%$ glycerin and 40$\%$ water, by weight) with a kinematic viscosity of $\nu_w=9.8 \times 10^{-6}$ m$^{2}$ s$^{-1}$, measured with a standard rheometer (Ares-G2, TA Instruments). Several $\Rey$ within a biologically relevant range ($\Rey = 645, 968,$ and $1291$) were tested by varying $f$. Across this range, the qualitative force trends remained consistent, with only modest changes in magnitude. For clarity, the results presented here focus on $\Rey = 968$, which represents the biologically relevant case.

The exopodite was designed to be neutrally buoyant within the glycerin-water mixture to eliminate the confounding effects of gravity on exopodite motions. The exopodite was secured to the root of the endopodite using a ball-bearing joint, allowing the exopodite to rotate passively in a plane parallel to the endopodite. We vary $\zeta$ from 0$^{\circ}$ to 80$^{\circ}$ in increments of 10$^{\circ}$, and include the biologically relevant $\zeta=35$\textdegree \ using a swappable ramal joint. Negative values of the $\zeta$ angle were not considered as they do not achieve pleopod abduction and adduction during the power and return strokes, respectively. The angle between the fixed endopodite and rotating exopodite, $\gamma$, passively actuated through fluid-structure interactions and was physically limited to a maximum of $\gamma=70$\textdegree \ in all experiments. Tracking of $\gamma$ was performed digitally using the DLTdv8 package for MATLAB \citep{Hedrick_2008}   from raw videos from a high-speed camera positioned directly anterior to the appendage, such that when the appendage reached the mid-stroke position, the endopodite was parallel to the focal plane. We used three reference points along the dotted lines on the endopodite and exopodite to estimate $\gamma$ (Fig S1).

The maximum value of $\gamma$ was chosen based on biological observations \citep{murphy2011metachronal}. To quantify the influence of cupping angle on kinematics, we computed two key metrics from the time-resolved $\gamma$ profiles. First, we calculated the peak abduction angular velocity during the power stroke, defined as $\Delta \gamma / \Delta t_{\text{power}}$, which captures how quickly the exopodite opens to its maximum angle. Rapid abduction is advantageous for thrust generation, as it increases the effective surface area early in the stroke. Second, we measured the fraction of the return stroke during which the exopodite remains abducted, using a threshold minimum $\gamma$ angle. This metric serves as a proxy for drag modulation, since maintaining a closed profile during the return stroke minimises drag. These parameters capture how effectively each cupping angle, $\zeta$, influences the passive kinematics of the abduction angle, $\gamma$.

\subsection{Flow Field Measurements}

\begin{figure}
    \centering
    \includegraphics[width=0.5\linewidth]{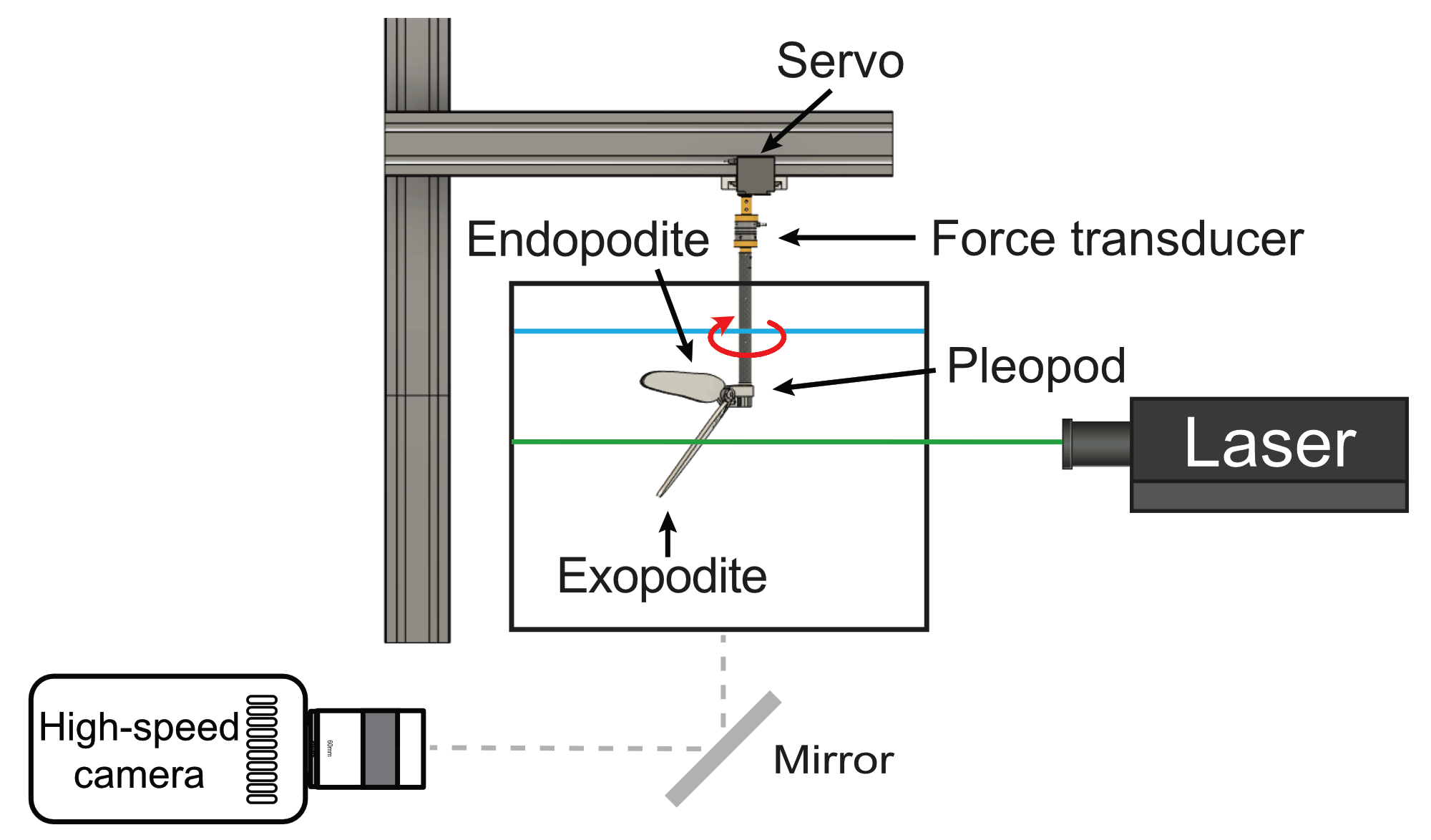}
    \caption{
        \raggedright
        \textbf{Schematics of the experimental setup for simultaneous force and PIV data collection.} The horizontal laser sheet bisected the fully abducted exopodite. Two lasers mounted opposite each other produced two coplanar overlapping laser sheets to eliminate shadows (the figure shows only one laser, for clarity). The pleopod was mounted horizontally and oscillated along the horizontal plane (red arrow)  by a servomotor. The force transducer was mounted above the waterline between the servo and the model using custom adapters, aligning the $x$-axis of the force transducer with the anterior face of the endopodite. Experiments were conducted in a 210-liter tank ($119\times43\times41$ cm$^3$), with an 11.5 cm long paddle, and a distance 1.2L to the nearest tank walls. Note that the experimental tank is not to scale.}
    \label{fig:experiments}
\end{figure}

Two-dimensional particle image velocimetry (PIV) was performed to characterize the near-flow field around the exopodite. The horizontal laser sheet bisected the exopodite paddle at maximum actuation ($\gamma=70^{\circ}$). Two continuous lasers formed overlapping laser sheets (4W at 532 nm, Optotronics, Mead, CO) to eliminate shadows on either side of the pleopod. A high-speed camera (NOVA-R3, Photron, Tokyo, Japan) recorded at 250 frames per second and a shutter speed of 1/500 s. A mirror angled at $45^{\circ}$ was mounted under the tank to record the bottom view (equivalent to the lateral view of the pleopod, figure \ref{fig:experiments}). The fluid was seeded with 10 $\mu$m neutrally buoyant hollow glass spheres (LaVision, Göttingen, Germany). The flow fields were analyzed using DaVis 10 software (LaVision, Göttingen, Germany), with three passes at 48 pixels $\times$ 48 pixels (50\% overlap) and three passes at 32 pixels $\times$ 32 pixels (75\% overlap). Post-processing used universal outlier detection, correlation-based vector rejection, median filtering, and interpolation of isolated missing vectors. To estimate the force contribution of leading-edge vortices (LEVs), circulation was computed from the PIV-resolved vorticity fields and used to estimate the force using a vortex impulse method \citep{siala2019leading}. The resulting impulse-based lift and thrust estimates were compared with direct force measurements to quantify the contribution of coherent vortex dynamics to the total measured forces.

\subsection{Force Measurements}
\begin{figure}
    \centering
    \includegraphics[width =0.75\linewidth]{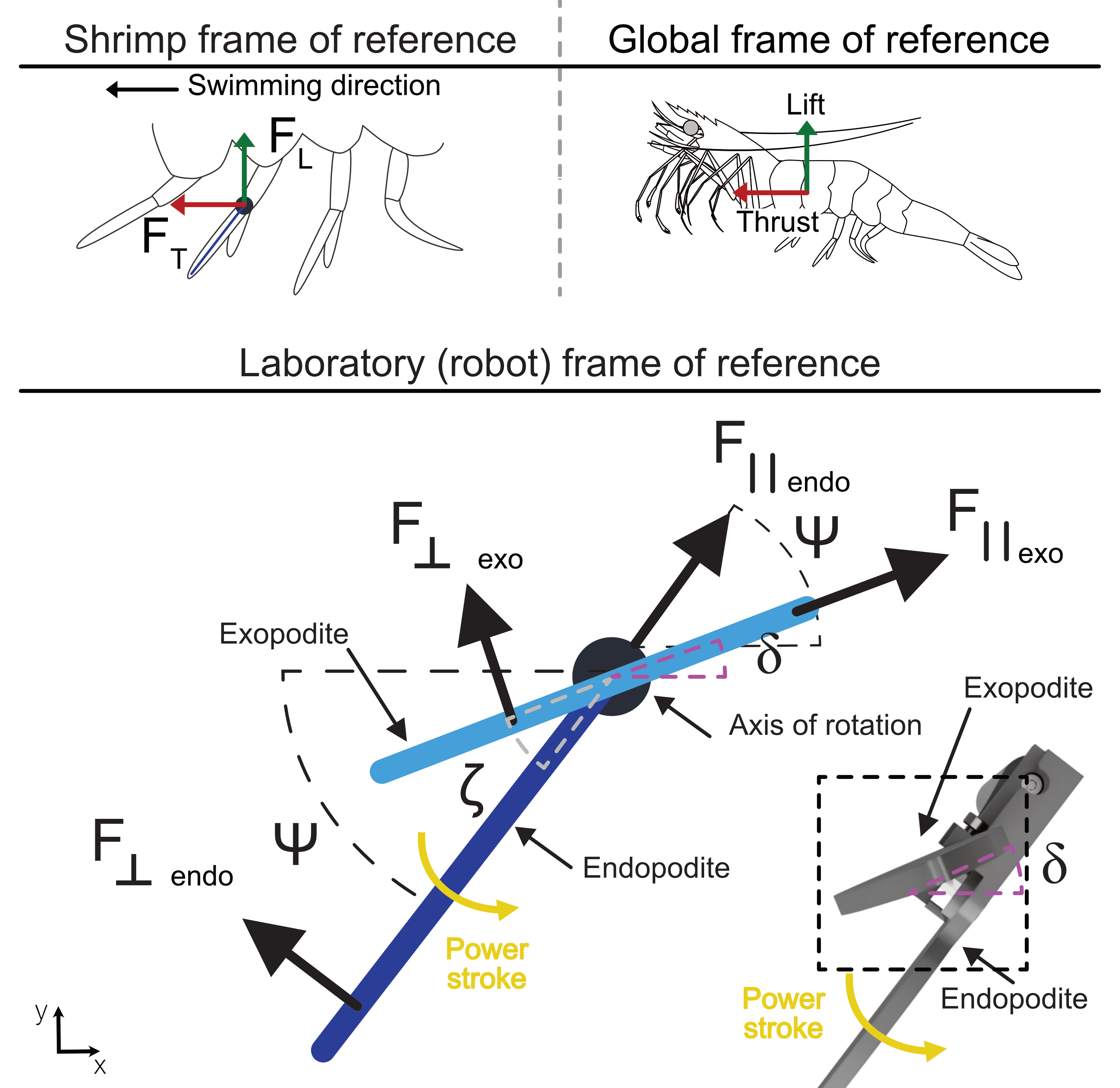}
    \caption{
            \raggedright
            \textbf{Force frame of reference}. The forces are shown in three frames of reference: laboratory, shrimp, and global frames. Forces on the rotating paddle (blue) are measured in two directions, one that is perpendicular ($F_{\perp}$) and one that is parallel to the endopodite and exopodite ($F_{\parallel}$). In the shrimp frame of reference, $F_{T}$ is the force in the swimming direction, and $F_{L}$ is the force orthogonal to the swimming direction. In the global frame of reference, lift is the force pointing vertically, and thrust is the horizontal force. $\psi$ is the angle between the body axis and the endopodite, $\zeta$ is the angle between the endopodite and exopodite, and $\delta$ is the angle between the exopodite and the horizontal direction. The paddle illustration shows the lateral view in the laboratory frame of reference, with the axis of rotation, endopodite, and exopodite labeled.}
    \label{fig:frameRef}
\end{figure}

Hydrodynamic forces were measured using a six-axis force transducer (Nano 17 F/T, ATI Industrial Automation, Rochester Hills, MI, USA) with a resolution of 3.1 mN, and a sampling rate of 100 Hz. The transducer was mounted between the servomotor and the pleopod model and above the waterline using custom adapters. The transducer was aligned such that the measured components corresponded to the pleopod frame of reference: the longitudinal force, $F_\parallel$, acting parallel to the pleopod model, and the orthogonal force $F_\perp$, acting normal to it (figure \ref{fig:frameRef}). 

Raw signals were denoised using multiscale principal component analysis (MSPCA) implemented in MATLAB (R2024a, Wavelet Toolbox; wmspca). A level-6 wavelet decomposition was performed using the Coiflet wavelet coif3 with symmetric boundary extension \citep{bakshi1998multiscale, aminghafari2006multivariate, daubechies1988orthonormal, daubechies1992ten}. At each scale, principal components were retained according to the npc='nodet' criterion (MATLAB wmspca default), and the denoised signals were reconstructed from the retained components. Data collection was initiated after a 60-second start-up period to ensure the flow had reached a periodic steady state. Forces were then recorded over four consecutive beat cycles and cycle-averaged.

%(figure \ref{fig:frameRef}).

%In the context of this work, the shrimp and the global frames of reference are the same. For each $\zeta$, the projected area was computed over the stroke cycle from the measured exopodite kinematics ($\gamma$) and the projected endopodite and exopodite profiles, with the endopodite reference orientation taken when it was perpendicular to the focal plane. The perpendicular force on each ramus was estimated as: $F_{\perp,i} = F_{\perp}A_i/A_{\text{tot}}$, where $i=\text{endo}$ or $\text{exo}$, $A_i$ is the projected area of the endopodite or exopodite, and $A_{\text{tot}} = A_{\text{endo}} + A_{\text{exo}}$ is the total projected area. 

The measured force components in the pleopod frame of reference, $F_\parallel$ and $F_\perp$, were transformed into the axial and normal components relative to the swimming direction in the shrimp frame of reference, $F_{T}$ (thrust) and $F_{L}$ (lift). It must be noted that the measured $F_\parallel$ and $F_\perp$ represent the combined contribution of the endopodite and the exopodite: $F_\parallel = F_{\parallel,\text{endo}} + F_{\parallel,\text{exo}}$ and $F_\perp = F_{\perp,\text{endo}} + F_{\perp,\text{exo}}$. 
To estimate individual contributions, $F_\perp$ was decomposed based on the projected areas of each rami onto the vertical plane normal to the swimming direction: $F_{\perp,i} = F_{\perp}A_i/A_{\text{tot}}$, where $i=\text{endo}$ or $\text{exo}$, $A_i$ is the projected area of the endopodite or exopodite, and $A_{\text{tot}} = A_{\text{endo}} + A_{\text{exo}}$ is the total projected area. The projected areas were computed over the stroke cycle from the measured exopodite kinematics ($\gamma$) and ramus profiles, for each $\zeta$, applying perspective and parallax corrections. Projected areas of the exopodite and endopodite were measured digitally when each appendage lay approximately parallel to the imaging focal plane.

As a first-order approximation, the parallel force $F_\parallel$ was divided equally between the rami based on their similar mass and geometry, as it represents the component acting parallel to their surface planes. Using this decomposition, the thrust and lift contributions from each ramus was computed using: 
\begin{equation}
    \begin{bmatrix}
        F_{L,i} \\
        F_{T,i}
    \end{bmatrix}
    =
    \begin{bmatrix}
        \cos\theta_i & \sin\theta_i \\
        \sin\theta_i & -\cos\theta_i
    \end{bmatrix}
    \begin{bmatrix}
        F_{\perp,i} \\
        F_{\parallel,i}
    \end{bmatrix},
    \label{eq:force_LT_endo}
\end{equation}
where $\theta_i$ is the angle of attack of each ramus: $\theta_i = \psi$ for the endopodite and $\theta_i = \delta$ for the exopodite.

The time-resolved hydrodynamic forces generated by the appendage were compared with predictions from a one-dimensional, reduced-order force model (described in the next section). In the lab reference frame, forces measured with the load cell were decomposed into the $F_{\perp}$ and $F_{\parallel}$ components (figure \ref{fig:frameRef}), which were then transformed into global-frame drag-based thrust and lift ($F_{T}$ and $F_{L}$) using equation \eqref{eq:force_LT_endo}. Forces were averaged across four consecutive beat cycles for each $\zeta$ angle.

\section{Analytical description of forces on the appendage}

An analytical description of the forces acting on the appendage was derived incorporating drag and added-mass, which are the dominant forces during appendage motion. The Basset history term was neglected due to its minimal contribution. The reduced-order model considers only the force component normal to the appendage surface, which corresponds to the experimentally measured orthogonal force $F_{\perp}$. This simplification is consistent with oscillating-plate systems in which the dominant hydrodynamic loading acts normal to the surface, while tangential contributions are comparatively small \citep{keulegan1958forces, shih1971drag, herrera2023omnidirectional}. We therefore model the total hydrodynamic force as the normal component $F_{\text{tot}} \equiv F_\perp$ and use it to evaluate the drag and added-mass contributions across a range of cupping angles $\zeta$. The reduced-order model is introduced as an independent estimate of ramus-specific loading, allowing us to assess when projected-area partitioning remains valid and when nonlinear vortex interactions may become important.

Assuming the setup effectively represents instantaneous translation, the total hydrodynamic force on the appendage can be described as the superposition of a drag force $F_{D}$ and  an inertial (added mass) force $F_{M}$: 
% \vspace{0cm}
\begin{equation}
F_\text{tot}\,= \frac{1}{2}\,\rho_w\,C_d\,A(t)\,v_{\perp}\,|v_{\perp}|
+ \rho_w\,C_m\,V\,\frac{\mathrm{d}v_{\perp}}{\mathrm{d}t}
\label{eq:OGmorison}
\end{equation}
% \vspace{0.2cm}
where $\rho_w$ is the fluid density, $v_{\perp}$ is the flow velocity normal to the appendage surface, $A$ is the time varying cross-sectional area, $C_d$ is the drag coefficient, $C_m$ is the inertia coefficient, and $V$ is the volume of fluid displaced by the body. We approximate the fluid velocity at the surface of the appendage by the velocity of the appendage itself. 

%Note that, to first approximation, we assume the variation in projected surface area during the stroke results in a negligible contribution from the time-dependence of the added mass force to the net hydrodynamic force.
%Because the projected area varies during the stroke, the effective added mass is time-dependent. However, the term proportional to the time derivative of this effective added mass is small compared with the drag and standard added-mass acceleration terms and is neglected.

In our experiment, the angular position of the appendage is a time-dependent function $\psi(t)$, defined to be zero when the appendage is parallel to the swimming direction. The translational velocity is given by $v(r,t) = r\dot{\psi}(t)$, whereby $r$ is the radial position along the appendage from the pivot point $O$. The drag force acting on a small segment of the appendage, $\Delta r$, with width $W$, is then expressed as
\begin{equation}
    \Delta F_D(r,t) = \frac{1}{2}\rho_w C_d (W \Delta r) r^2 \dot{\psi}(t) \left| \dot{\psi}(t) \right| ,
\end{equation}
which is integrated over the length of the appendage, $l$, to obtain the total drag force: 
\begin{equation}
    F_D(t) = \frac{1}{6}\rho_w C_d W l^3 \dot{\psi}(t) \left| \dot{\psi}(t) \right|.
\end{equation}
Similarly, the inertial (added mass) force acting on the same segment is
\begin{equation}
    \Delta F_M(t) = \rho_w C_m  V_p \frac{\mathrm{d}v}{\mathrm{dt}} = \rho_w C_m(W^2\Delta r)(r\ddot{\psi}),
\end{equation}
where $V_p$ is the displaced fluid volume, approximated as $\sim W^2\Delta r$, consistent with the added mass for a flat plate. Integrating over the length of the appendage yields the total inertial force:
\begin{equation}
    F_M(t) = \frac{1}{2}\rho_w C_m W^2 l^2 \ddot{\psi}(t).
\end{equation}
Thus, the total force on the appendage is given by 
\begin{equation} \label{eq:totalForceMorisonArea}
    F_{\text{tot},i}(t) = F_{D,i}(t)+F_{M,i}(t) = 
    \frac{1}{6}\rho_w C_d A_i (\gamma,\zeta) l_i^2 \dot{\psi}(t) \left| \dot{\psi}(t) \right| + 
    \frac{1}{2}\rho_w C_m W_i^2 l_i^2 \ddot{\psi}(t),
\end{equation}
where $i=\text{endo}$ and $\text{exo}$, and $A_i$ is the time-varying projected area as a function of cupping angle $\zeta$ and the abduction angle $\gamma$. The time-varying projected area was computed from the measured exopodite abduction angle $\gamma$ using the geometric procedure described in in section 2.3 (see Supplementary figure 1).

Here, we use a first approximation of the drag and inertia coefficients. Geometry-specific drag coefficients for rotating plates at the present aspect ratio and kinematic regime are not available in the literature, precluding the use of empirically derived coefficients. We estimate the cycle-averaged drag and inertia coefficients using the Stokes-Wang analytical solution for oscillatory viscous flow \citep{stokes1851effect, wang1968high, sarpkaya1986force}:
\begin{align}
    C_{d,i} &= \frac{3 \pi^3}{2KC_i} \left[ (\pi \beta_i)^{-1/2} + (\pi \beta_i)^{-1} - \frac{1}{4} (\pi \beta_i)^{-3/2} \right],\\
    C_{m,i} &= 2 + 4 (\pi \beta_i)^{-1/2} + (\pi \beta_i)^{-3/2},
\end{align}
where $\beta_i = \Rey_i/KC_i$ and $KC_i=\psi_m l_i/W_i$ is the Keulegan-Carpenter number \citep{morison1950force, keulegan1958forces}. This amplitude-based expression is equivalent to the standard definition $KC = U_m T/D$, written here in terms of the appendage stroke amplitude $\psi_m l_i$ and characteristic width $W_i$. Note that these expressions have been shown to agree well with experimental and numerical data for small $KC$ $(\ll 1)$ and $\beta \gg 1$ \citep{sarpkaya1986force, justesen1991numerical}. At higher $KC$ (typically between 0.7 and 1), vortex shedding and vortical instabilities cause deviations from the Stokes-Wang predictions for $C_d$ and $C_m$ \citep{sarpkaya1986force}. We computed $\Rey$ and $KC$ for both the endopodite and exopodite and used them to estimate $C_d$ and $C_m$ for each component, from which we calculated the total force $F_{\text{tot},i}$. 
Finally, the lift and thrust were estimated using \eqref{eq:force_LT_endo}, where the perpendicular force  $F_{\perp,i}$ is taken as the total force $F_{\text{tot},i}$ and the parallel force is zero, $F_{\parallel,i} = 0$. This follows from the reduced-order model, which approximates the local flow velocity using the appendage kinematics and therefore resolves only the velocity component normal to the appendage surface.

% The dominant forces acting on the appendage are the drag force and the added mass force. Secondary forces, including the Basset force, as well as centrifugal and Coriolis contributions, are neglected for simplicity. We reduce the system to a one-dimensional model that resolves the component of force aligned with the local velocity vector of the appendage. This simplification occurs through a shift in the reference frame from the global coordinate system to the appendage local reference frame (fig. \ref{fig:frameRef}). 

% Similar approaches have been employed in previous studies of oscillatory flows and bio-inspired locomotion, where the stream-wise force component accounts for the majority of the hydrodynamic loading \citep{keulegan1958forces, shih1971drag, herrera2023omnidirectional}. While this model neglects transverse forces, it provides a method to estimate the relative contributions of drag and inertial effects across a range of motion parameters, particularly in the context of varying Keulegan–Carpenter numbers ($KC$). This one-dimensional model is best adapted for $KC \leq 5$, where the flow is inertia-dominated, and $KC$$ \geq 25$, where the flow is drag-dominated \citep{hogben1977estimation}. At intermediate $KC$, the flow is complex, transitioning from drag to inertia-dominated forces. 

\section{Results}
\subsection{Role of the cupping angle in driving appendage abduction}
The projected surface area of the pleopod analog can be readily modulated by varying the abduction angle $\gamma$, the orientation of the cupping angle $\zeta$ with respect to the endopodite, and the net overlap of the exopodite and endopodite. By optimizing the pleopod surface area during a stroke, $\gamma$ plays a key role in the hydrodynamic performance of the appendage: it reaches its maximum during the power stroke to magnify thrust and is zero during the return stroke to minimize drag. In addition to modulating the effective surface area, the actuation of $\gamma$ also enables the extension of the exopodite into the free stream to modulate thrust and lift via changes in the orientation of the induced forces and the position and stability of particular flow features like LEVs.

\begin{figure}
    \centering
    \includegraphics[width=1\linewidth]{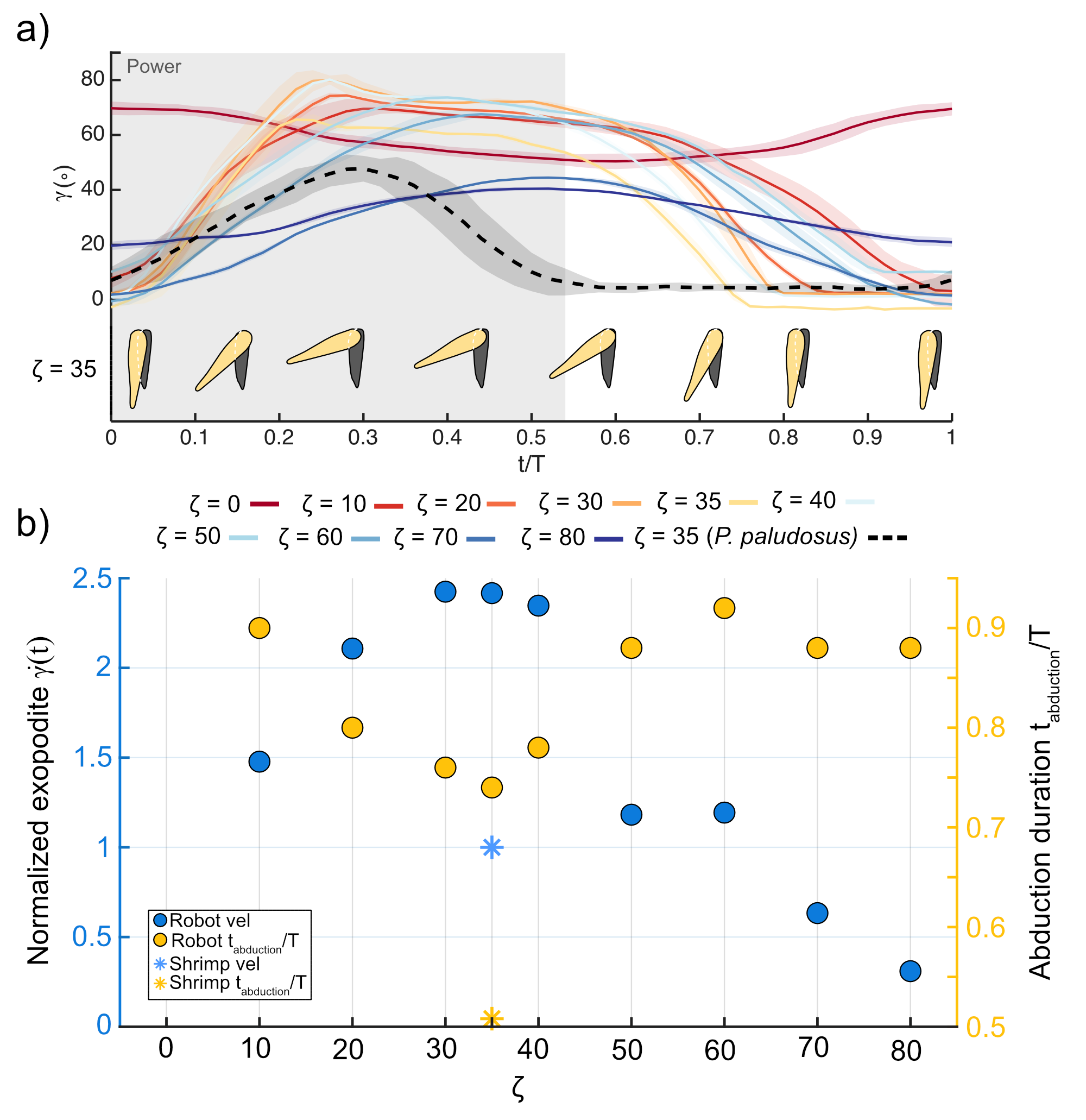}
    \caption{
    \raggedright
    \textbf{Variation of $\gamma$ throughout one beat cycle.} a) Phased-average exopodite kinematics during a beat at varying cupping angles and $\Rey = 968$. The instantaneous time $t$ during the beat cycle is normalized to the beat period $T$. Shading indicates the standard deviation from the plotted mean values (solid lines) of three consecutive beats. The dashed black line represents biological observations obtained from the kinematics of tethered \emph{P. paludosus}. Profiles are drawn for $\zeta = 35^o$, where the exopodite is shaded yellow.  b) Normalized angular velocity ($\dot{\psi}(t)$) to maximum exopodite angle and abduction duration (proportion of the cycle during which the exopodite is abducted). Angular velocity is normalized to the shrimp value (*). Filled circles     
  represent experimental data; asterisks represent biological shrimp data at $\zeta = 35^o$}
    \label{fig:GammaOverTime}
\end{figure}

Both the rate and the amplitude of abduction and adduction are key to generating sufficient thrust and lift over a cycle (normalized time, $t/T$) and are highly dependent on $\zeta$ (figure \ref{fig:GammaOverTime}). Due to this coupling, $\zeta$ directly influences the dynamics of the total surface area change of the pleopod and its effective projected surface area to the flow throughout a cycle. For all cupping angles except $\zeta = 0^{\circ}$ and $80^{\circ}$, increasing the abduction angle led to an increase in the maximum effective profile area, particularly during the power stroke when $\gamma$ peaked. Conversely, the effective pleopod area was minimised during the return stroke. Although the wetted surface area was constant across all cases, the orthographic projection of the exopodite varied with the cupping angle. At the extremes, $\zeta = 0^{\circ}$ and $\zeta = 80^{\circ}$, the variation in $\gamma$ was smaller than for all the other cases because the lateral component of the force from hydrodynamic loading of the exopodite during the power stroke was the lowest due to the reduced exopodite projected area (thickness of the paddle). Kinematic analyses show that moderate cupping angles, $20^{\circ}$ \ $\leq \zeta \leq $ $40^{\circ}$, resulted in the fastest and earliest onset of exopodite abduction and adduction (figure \ref{fig:GammaOverTime}b). 

In the appendage model, peak exopodite abduction rate ($\Delta \gamma / \Delta t_{power}$) exhibited a non-monotonic dependence on cupping angle, with maximum rotational velocity occurring at $\zeta = 30^{\circ}$ (figure \ref{fig:GammaOverTime}b). Abduction and adduction speeds are lowest at both low and high cupping angles, reaching their lowest values at the smallest ($\zeta = 0^{\circ}$) and largest ($\zeta = 80^{\circ}$) cupping angles. The fraction of the return stroke when $\gamma$ was minimised showed similar trends. The exopodite adducted earliest at $\zeta = 35^{\circ}$, approximately midway through the return stroke (t/T = 0.74). 

The overall $\gamma$ dynamics of the robotic model were consistent with biological data. $\gamma$ increased during the power stroke, in particular during the acceleration phase of the pleopod and reached a minimum during the return stroke (figure \ref{fig:GammaOverTime}). The similarity in magnitude and timing of $\gamma$ during the power stroke between our passively actuated model and live \textit{P. paludosus} suggests that fluid-structure interactions primarily govern exopodite abduction. Because time-resolved $\gamma$ measurements are not available for \textit{P. vulgaris} (the species on which our robot is based), we compare our results to reference kinematics from \textit{P. paludosus}. Despite the species difference, the pleopod morphology and swimming kinematics are qualitatively similar between these two species. Adduction occurs much earlier in \textit{P. paludosus}, beginning prior to the onset of the return stroke. In contrast, exopodite adduction in our robotic model was consistently delayed past the power-to-return stroke transition. For $\zeta = 35^{\circ}$, the proportion of the return stroke during which the exopodite remained abducted was 1.4 times longer in our model than in \textit{P. paludosus} (figure \ref{fig:GammaOverTime}b), suggesting that active muscular control may accelerate adduction in live shrimp. To our knowledge, these measurements provide the first time-resolved characterization of exopodite abduction dynamics across cupping angles, enabling direct assessment of passive versus active actuation mechanisms.

\subsection{Appendage thrust and lift production}
\begin{figure}
    \centering
    \includegraphics[width=1\linewidth]{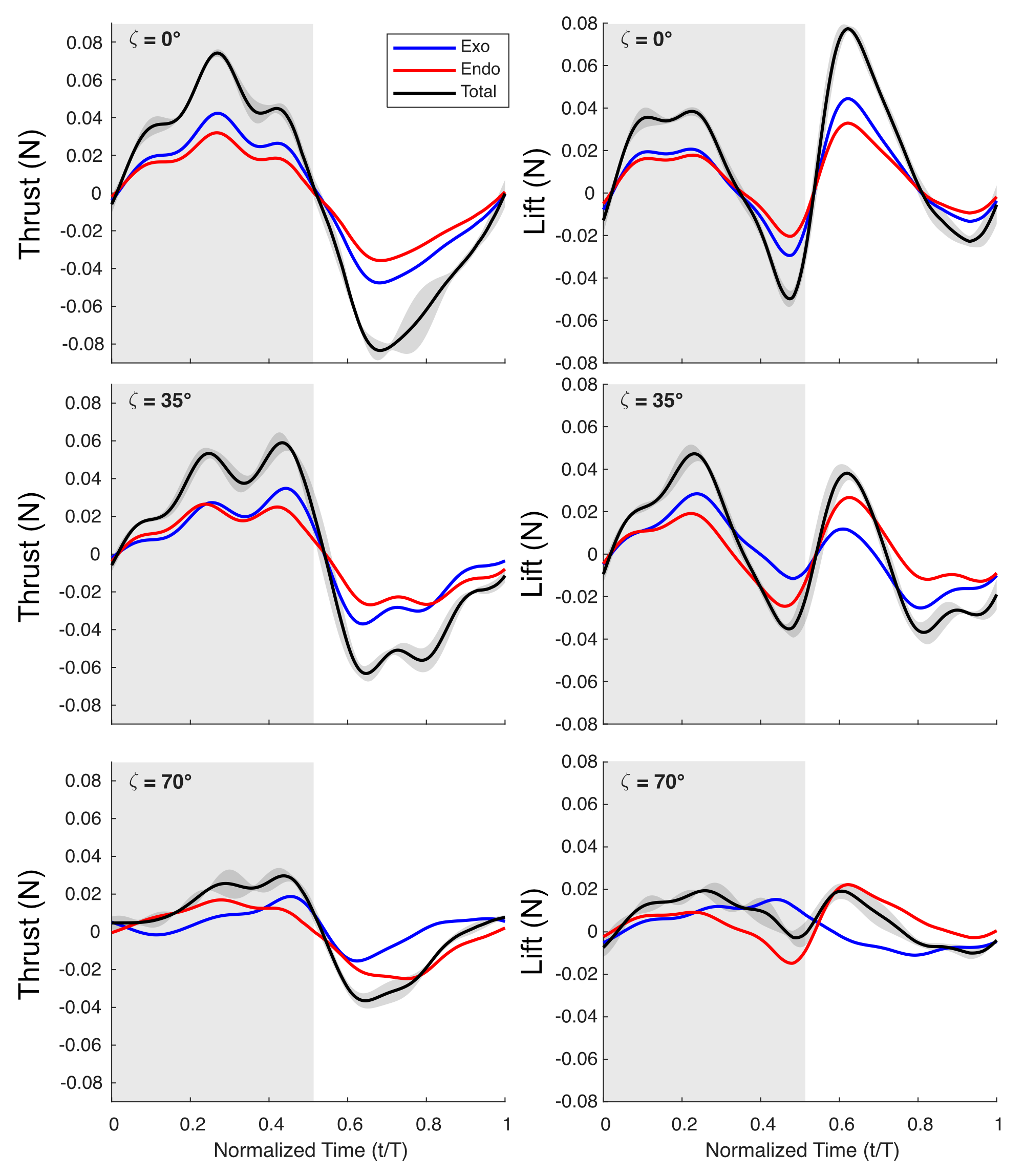}
    \caption{
    \raggedright
    \textbf{Dynamics of thrust and lift forces at variable cupping angles.} Values for the top, middle, and bottom panels are: $\zeta = 0^\circ$, $\zeta = 35^\circ$, and $\zeta = 70^\circ$, respectively. The total measured hydrodynamic force is shown with a black line (with shaded standard deviation (STD)). The decomposed contributions from the exopodite and the endopodite are shown in blue and red lines, respectively. Time $t$ is normalised using the stroke period $T$.}
    \label{fig:Forces_tT}
\end{figure}

Thrust and lift vary significantly throughout a beat cycle, with distinct patterns between the power and return strokes. During the acceleration phase of the power stroke ($0 < t/T < 0.25$), drag-based thrust increases sharply and reaches a positive maximum near $t/T \approx 0.25$ (figure \ref{fig:Forces_tT}). Thrust then decreases rapidly during the deceleration phase ($0.25 < t/T < 0.53$), approaching zero as the exopodite becomes fully abducted at the onset of the return stroke. During the return stroke ($0.53 < t/T < 1$), thrust reverses sign,  with the peak negative thrust (net drag) occurring during the initial acceleration phase of the recovering pleopod. While the overall dynamics of thrust and lift production are primarily governed by the oscillatory motions of the pleopod relative to the swimming direction, the surface area change and orientation of the exopodite motions modulate the overall magnitude of and the dynamics between thrust and lift. %These dynamics reflect the changing surface area and orientation of the appendage rami throughout the cycle.

Based on the projected areas of the endopodite and the exopodite and their measured kinematics, we decompose the total hydrodynamic forces into separate contributions from each branch across a normalised stroke cycle (t/T) for three cupping angles: $\zeta = 0^\circ$, $35^\circ$, and $70^\circ$ (figure \ref{fig:Forces_tT}). These three specific values for $\zeta$ are relevant given that $\zeta = 0^\circ$ is the baseline case without cupping; $\zeta = 35^\circ$ represents a case where $\gamma$ reaches its maximum value and is also the biologically relevant angle ($\zeta \sim 37^\circ$ measured in \textit{Euphausia superba}, \cite{oliveira2023pleobot}); and $\zeta = 70^\circ$ is the largest cupping angle at which the exopodite still returns to $\gamma \sim 0^\circ$ during the return stroke.

For the first study case, at $\zeta = 0^\circ$, the individual force contributions of the exopodite and the endopodite are in phase, since the two structures are coplanar (difference of 0.001 t/T). This results in constructive forces that enhance the total thrust and lift throughout the cycle. At $\zeta = 35^\circ$, phase alignment between the exopodite and endopodite begins to diverge (average differences of 0.021 t/T for thrust and 0.0195 t/T for lift), and the exopodite contribution exceeds that of the endopodite. These trends indicate the emergence of functional differentiation between rami, with the exopodite increasingly acting as a lift-enhancing surface while the endopodite remaining primarily thrust-dominated. 

At $\zeta = 70^\circ$, the thrust and lift contributions are completely out of phase (difference of 0.175 t/T for drag and 0.04 t/T for lift with opposite signs). In this case, the local lift peak produced by the exopodite coincides with a minimum contribution from the endopodite (figure \ref{fig:Forces_tT}, bottom panel), leading to partial cancellation of instantaneous forces and a reduction in total peak thrust and lift relative to intermediate $\zeta$. In contrast, the relative dominance of the two rami at $\zeta = 70^\circ$ shifts between the acceleration and deceleration phases of the power stroke, where the endopodite contributes more strongly early in the stroke, and the exopodite becomes the primary force-producing surface later in the stroke. This demonstrates that appendage substructure plays a critical role in modulating force production and that there exists a range of cupping angles for optimal thrust and lift generation.

\begin{figure}
    \centering
    \includegraphics[width=1\linewidth]{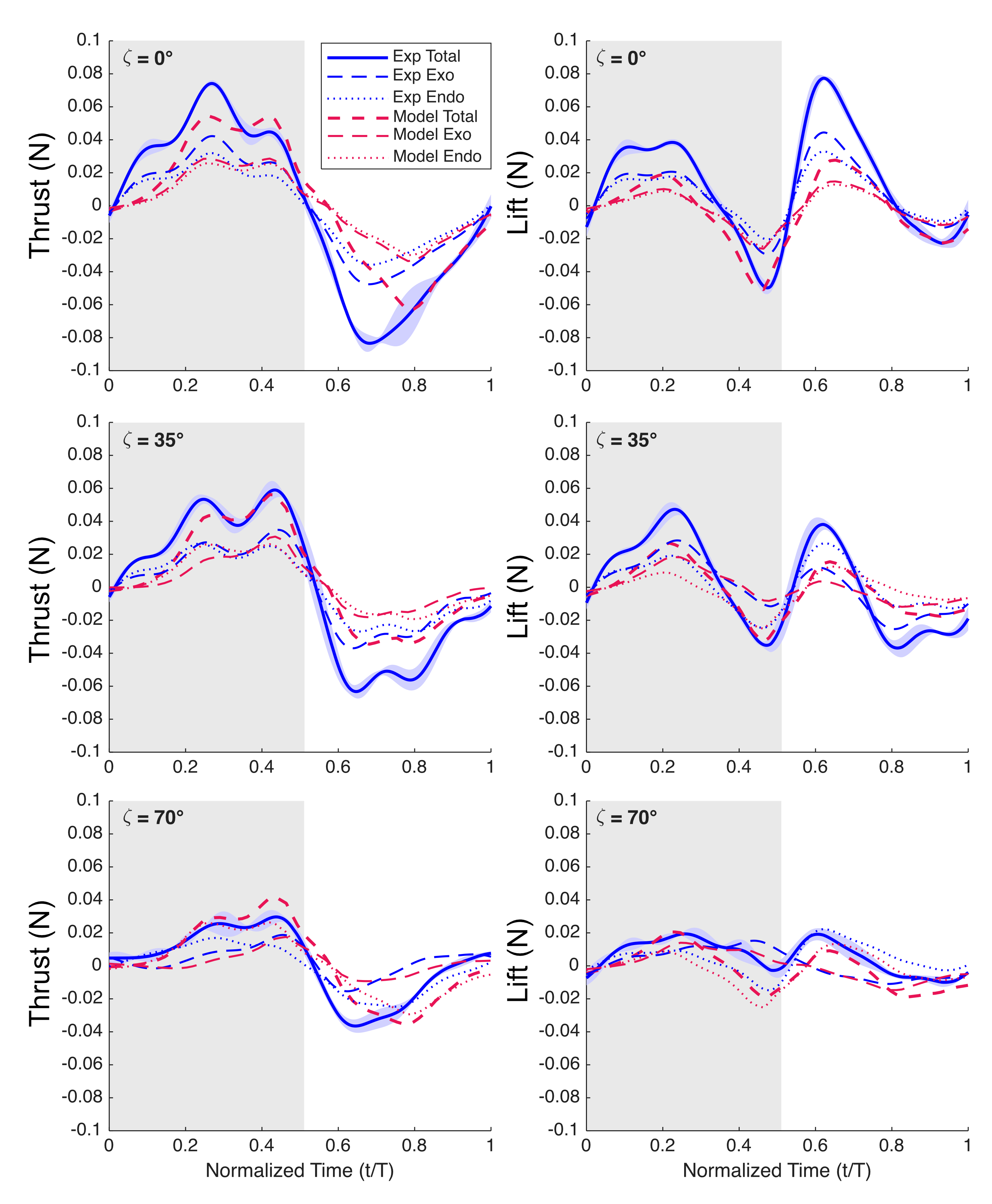}
    \caption{
    \raggedright
    \textbf{Experimental and model-predicted forces}. Comparison between experimentally measured forces (blue lines) and theoretical predictions (red lines) for thrust (left) and lift (right) at $\zeta = 0^\circ$ (top), $\zeta = 35^\circ$, and $\zeta = 70^\circ$ (bottom). Theoretical forces were estimated using \ref{eq:totalForceMorisonArea}. Forces were estimated for the endopodite and exopodite, with the latter including $\zeta$. The shaded blue area represents $\pm1$ standard deviation (n = 4 consecutive beats). Time was non-dimensional as $t/T$.}
    \label{fig:LiftDragComparison_zeta35}
\end{figure}

We compared the experimental time-resolved thrust and lift forces with those predicted by our reduced-order model to further evaluate the consistency between measured forces and the trends implied by the model assumptions (figure \ref{fig:LiftDragComparison_zeta35}). This model estimates force contributions from the exopodite and endopodite independently and reconstructs the total force as their linear sum. For all three $\zeta$ values, the model successfully qualitatively predicts the force trends, including the timing of peak thrust during the power stroke ($t/T \approx 0.35$) and lift generation during the return stroke ($t/T \approx 0.6$). At $\zeta = 0^\circ$, the model closely captures both amplitude and phase of the exopodite and endopodite forces over the whole cycle (NRMSE = 13.0 \%, $R^2 = 0.830$). At $\zeta = 35^\circ$, the model is consistent with the thrust peak measured during the power stroke near $t/T = 0.45$. Peak magnitude is closely matched (56.1 mN predicted versus 59.0 mN measured; peak ratio = 1.05), with minimal phase offset (phase shift = 0.014 $t/T$). During the return stroke ($0.53 < t/T < 1$), the phase and trend of the drag curve are relatively consistent, although the predicted forces slightly underestimate the experimental thrust reversal (NRMSE = 12.7 \%, $R^2 = 0.857$). At $\zeta = 70^\circ$, the model continues to reproduce the overall force trends, although discrepancies increase relative to lower cupping angles. For thrust, the model overpredicts the peak magnitude (41.0 mN predicted versus 29.7 mN measured; peak ratio = 0.72) while maintaining reasonable phase agreement (phase shift = 0.016 $t/T$; NRMSE = 14.2 \%, $R^2 = 0.791$).

Agreement between experimental measurements and model predictions indicates that the unsteady force dynamics introduced by moderate cupping can be captured using a reduced-order, kinematics-based framework in which the endopodite and exopodite forces are treated as approximately additive. The experimental measurements provide validation of the model assumptions, while the model offers an independent means of interpreting the force decomposition and predicting how changes in cupping angle and appendage motion change thrust–lift balance. This supports the applicability of equation~\ref{eq:totalForceMorisonArea} when extended to account for time-varying projected area and rotational effects.

\begin{figure}
    \centering
    \includegraphics[width=0.9\linewidth]{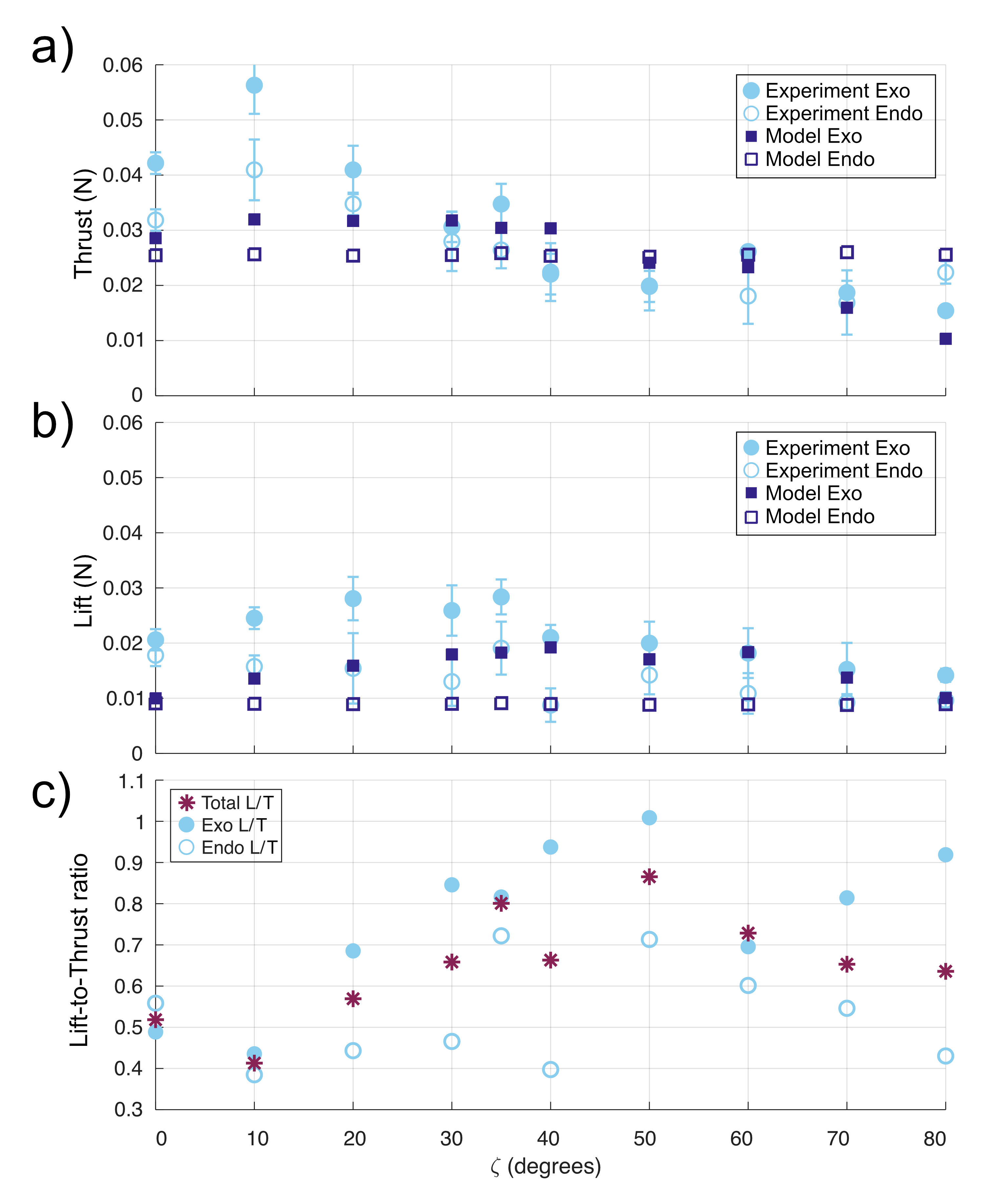}
   \caption{
\raggedright
\textbf{Thrust, lift, and lift-to-thrust ratio across cupping angles ($\zeta$) in the shrimp frame of reference.}
(a) Peak thrust during the power stroke, decomposed into exopodite and endopodite contributions from experimental measurements (blue circles) and model predictions (squares). (b) Peak lift during the power stroke, showing the corresponding experimental and model component contributions. (c) Experimental lift-to-thrust ratio ($L/T$), computed from the peak forces for the total appendage, exopodite, and endopodite. The total $L/T$ increases from low cupping angles to a maximum near $\zeta = 50^\circ$, indicating that pleopod kinematics can achieve comparable lift and thrust magnitudes at intermediate $\zeta$. At larger cupping angles, the total ratio remains moderate ($L/T \gtrsim 0.6$), while the absolute force magnitudes decrease.
}
    \label{fig:3panels}
\end{figure}

The stroke-averaged peak forces and lift-to-thrust ($L/T$) ratios vary systematically across the cupping angle sweep ($\zeta = 0^\circ$ - $80^\circ$). Thrust decreases monotonically with increasing $\zeta$ after $\zeta = 10^\circ$ (figure \ref{fig:3panels},a), while lift shows a non-monotonic trend, peaking at intermediate angles ($\zeta = 30^\circ$ - $40^\circ$) (figure \ref{fig:3panels},b). Experimental peak thrust is highest at $\zeta = 10^\circ$ and steadily declines with increasing $\zeta$. The decomposition by ramus shows that the exopodite contributes most significantly to lift, with a high of 62 \% of lift at $\zeta = 35^\circ$. It also contributes the majority of drag-based thrust between $\zeta = 0^\circ$ and $\zeta = 60^\circ$, between 50.7 \% and 58.8 \%. Within this range, the thrust produced by the endopodite remains relatively constant. The decline in exopodite thrust with increasing $\zeta$ arises from geometric changes in the effective area presented to the flow (figure \ref{fig:3panels},a). As the cupping angle increases, a smaller fraction of the exopodite surface remains orthogonal to the thrust-producing direction during the power stroke, reducing the projected area normal to the incident flow and therefore diminishing drag-based thrust generation (Supplementary figure 2).

The lift forces in figure \ref{fig:3panels},b span a narrower range than thrust across the cupping-angle sweep ($F_L \approx 0.009$–$0.03$ N). Both experiments and model predictions show that lift is maximized at intermediate cupping angles, with a peak occurring between $\zeta = 20^\circ$ and $50^\circ$. In this regime, the exopodite contributes most strongly to lift, consistent with the development of coherent leading-edge vortices (LEVs) observed in the vorticity fields (figure \ref{fig:vorticityFields}). At lower cupping angles, however, the endopodite contributes a substantial fraction of the total lift and, at $\zeta = 0^\circ$, produces lift comparable to that of the exopodite. This shows that lift generation is not exclusively associated with exopodite vortex formation, but instead is a result of contributions from both rami, with their relative importance shifting across the cupping-angle sweep.

The relative balance between lift and thrust varies systematically with $\zeta$, as quantified by the lift-to-thrust ratio in figure \ref{fig:3panels}(c). At low cupping angles ($\zeta \leq 10^\circ$), the total $L/T$ remains close to $\sim 0.5$, indicating thrust-dominated force production. As $\zeta$ increases, $L/T$ rises steadily and reaches a maximum near $\zeta = 50^\circ$, where lift and thrust become comparable in magnitude. This demonstrates that pleopod morphology and kinematics are capable of producing as much lift as thrust. However, this is not necessarily ideal for forward swimming. A moderate $L/T$ is therefore likely to be preferred, providing sufficient lift while maintaining strong net thrust production. At higher cupping angles ($\zeta \geq 60^\circ$), the total $L/T$ remains above $\sim 0.6$, but the absolute forces decline, suggesting diminished hydrodynamic effectiveness despite the maintained ratio.

The non-monotonic behavior around $\zeta = 40^\circ$ arises primarily from a reduction in the endopodite contribution, whereas the exopodite continues to exhibit elevated $L/T$ values. This variability suggests that the endopodite force production is more sensitive to changes in cupping angle and coordination than initially expected, potentially reflecting partial shielding or altered timing between the rami.

The prevalence of intermediate cupping angles in biological swimmers ($\zeta \approx 35^\circ$) therefore suggests that shrimp operate near a compromise regime, where force magnitudes remain high while maintaining a moderate lift-to-thrust balance ($L/T \approx 0.7$–$0.8$). At this angle, the exopodite can abduct rapidly and remain extended into the free stream for a larger fraction of the stroke, promoting LEV formation and efficient propulsion. The consistently higher exopodite $L/T$ across $\zeta$ further supports the idea of functional specialization between the rami.

\subsection{Contribution of the leading-edge vortex to lift}

The leading-edge vortex that forms on the exopodite during the power stroke is a critical flow structure responsible for generating lift. As the exopodite rapidly abducts and rotates into the flow, it creates conditions favorable for vortex formation along the leading edge. Similar unsteady mechanisms are well-documented in flapping flight, where LEVs enable high lift coefficients \citep{tian2006direct}. In shrimp, the ability to exploit an LEV may help offset negative buoyancy and maintain vertical positioning in the water column. Given the transient and geometry-dependent nature of LEVs, we quantify their contribution to force production by comparing impulse-force derived lift and thrust estimates from PIV with direct force measurements. This analysis helps assess how much of the total force can be attributed to the LEV and how its contribution varies with cupping angle $\zeta$.

\begin{figure}
    \centering
    \includegraphics[width=0.9\linewidth]{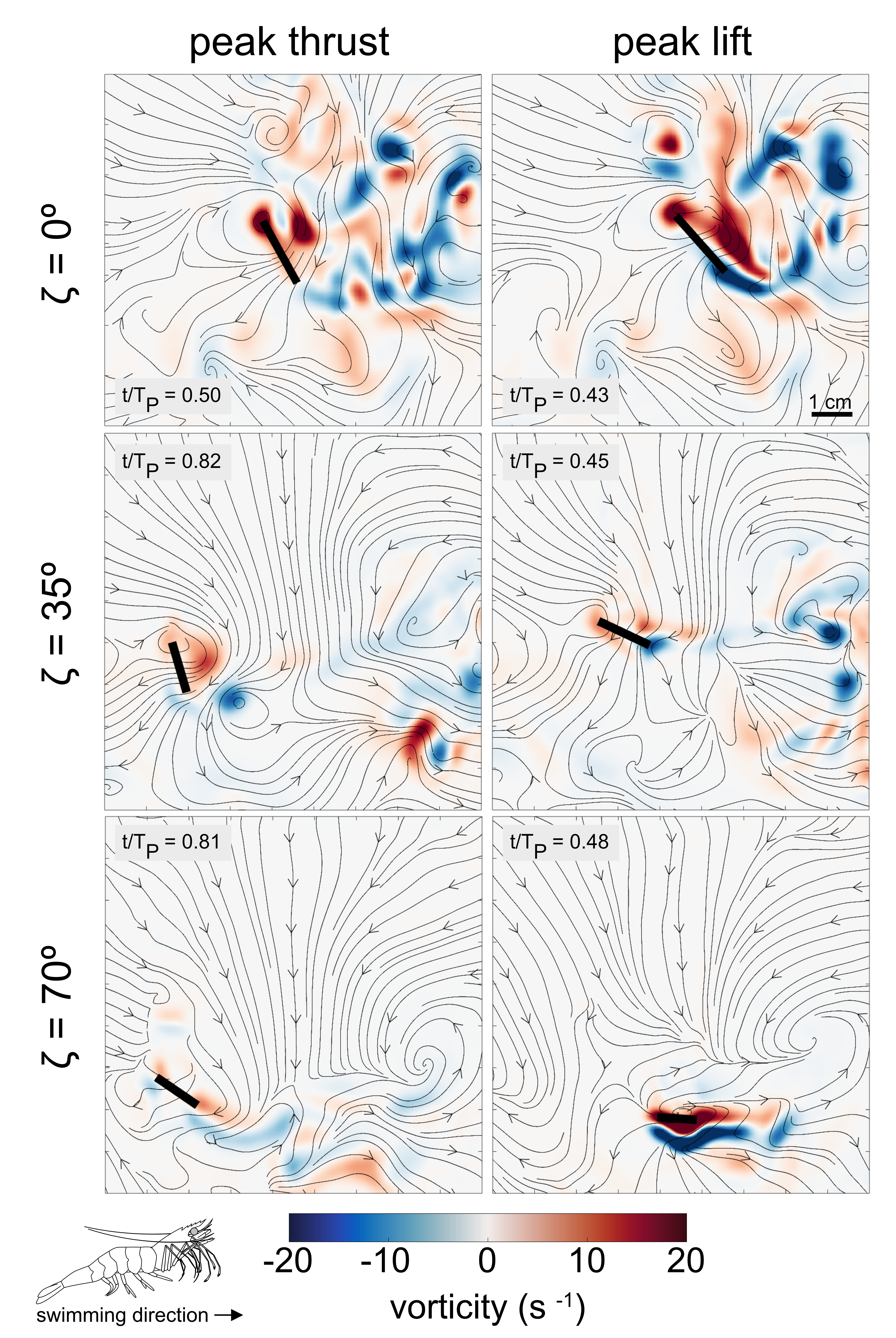}
    \caption{
        \raggedright
        \textbf{Vorticity fields and streamlines}. Chordwise vorticity fields and streamlines at maximum measured thrust (left column) and maximum measured lift (right column) for three cupping angles: $\zeta = 0^\circ$, $35^\circ$, and $70^\circ$ (top to bottom). Moderate cupping angles show strong and coherent leading-edge vortex (LEV) formation during the power stroke, while extreme $\zeta$ values exhibit weaker or detached vortex structures. The colour bar represents the vorticity magnitude for all plots. The black profile in each plot represents the cross-section of the exopodite where the laser sheet impinged. The raw PIV images are provided in the Supplementary Information Figure 3.}
    \label{fig:vorticityFields}
\end{figure}

Chordwise vorticity fields during the power stroke illustrate differences in vortex structure across $\zeta$ (figure \ref{fig:vorticityFields}). At low cupping angles ($\zeta = 0^\circ$), regions of elevated vorticity are present, but they are largely distributed within the wake and influenced by residual circulation from the previous stroke. The vorticity does not organize into a clearly bounded, persistently attached leading-edge structure along the exopodite. The corresponding thrust impulse magnitude derived from the vortex field is $1.07 \times 10^{-3}$ N·s, while the lift impulse magnitude is $3.88 \times 10^{-4}$ N·s. At intermediate angles ($\zeta = 35^\circ$), although peak vorticity magnitude is lower than at $\zeta = 0^\circ$, the flow exhibits a more coherent and spatially localized vortex, attached near the leading edge of the exopodite. Streamlines wrap around the appendage, indicating organized circulation that remains associated with the surface over a substantial portion of the power stroke. Consistent with this organization, the thrust impulse magnitude increases to $1.22 \times 10^{-3}$ N·s, the largest among the cases described here, and the lift impulse magnitude reaches $6.15 \times 10^{-4}$ N·s. At high cupping angles ($\zeta = 70^\circ$), vortex formation is more transient. While vorticity is generated near the leading edge, the LEV becomes unstable and detaches early, reducing near-field circulation and diminishing coherent vortex structure. The corresponding thrust and lift impulse magnitudes decrease to $3.27 \times 10^{-4}$ N·s and $1.95 \times 10^{-4}$ N·s, respectively. These observations suggest that intermediate cupping angles provide the most favorable conditions for persistent LEV development. The sign of the impulse depends on circulation orientation and its projection into the shrimp frame of reference; here, impulse magnitudes are reported to compare vortex-induced loading across $\zeta$.

\begin{figure}
    \centering
    \includegraphics[width=1\linewidth]{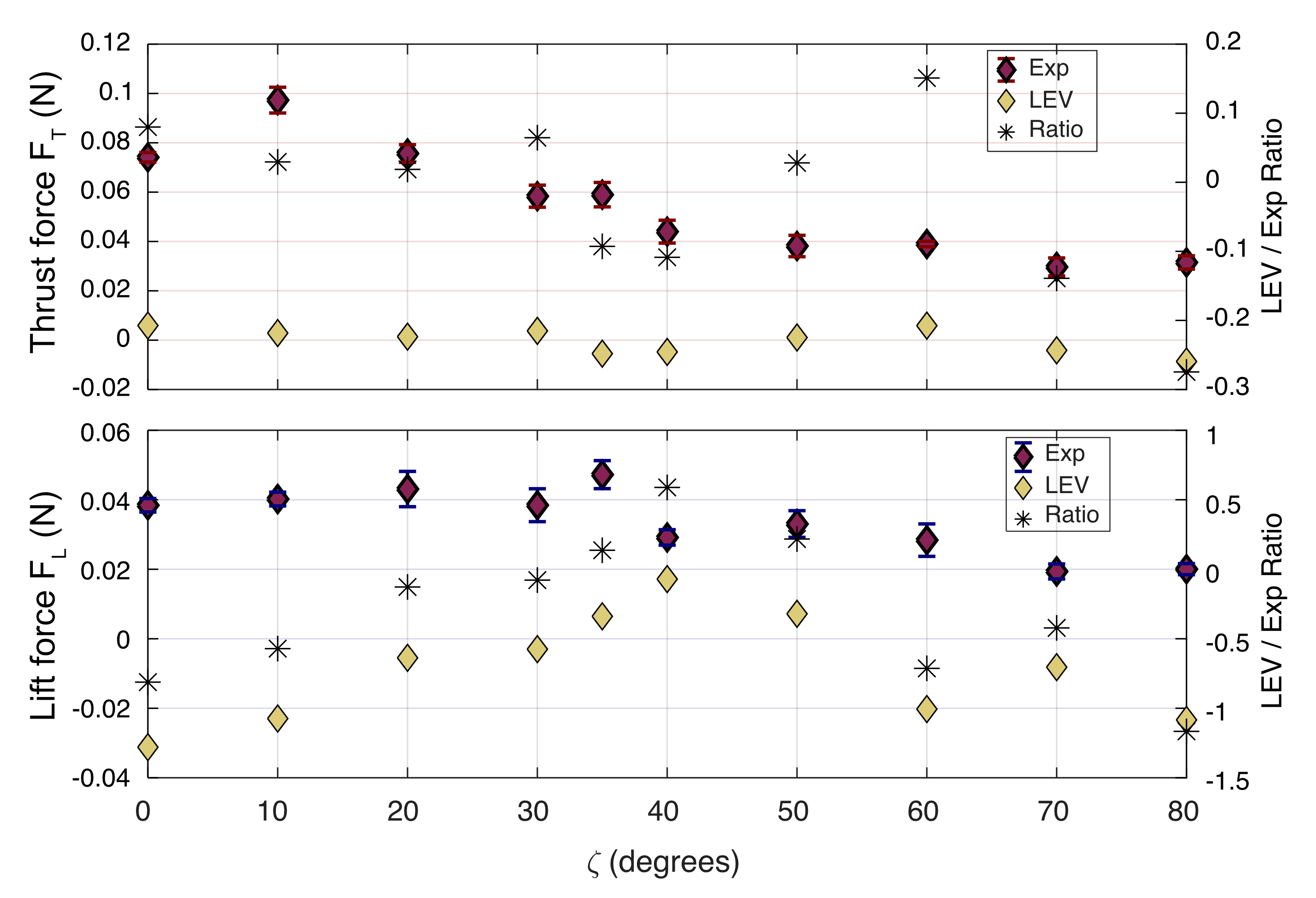}
    \caption{
        \raggedright
        \textbf{Comparison of LEV-derived phase-averaged total forces from PIV with experimental force measurements across cupping angles ($\zeta$)}. (a) Thrust force component ($F_T$): experimentally measured values (dark red diamonds), LEV-derived values from PIV (yellow diamonds), and their  ratio (LEV/Exp; asterix, right axis). (b) Lift force component ($F_L$): experimentally measured values (dark red diamonds), LEV-derived values (yellow diamonds), and LEV/Exp ratio (asterix, right axis). Ratios quantify the relative contribution of the LEV to the measured force component. Positive values indicate that the LEV contribution projects in the same direction as the measured force, whereas negative values indicate an opposing contribution, reducing the net thrust or lift. The largest positive lift contributions occur at intermediate cupping angles, consistent with coherent LEV attachment (figure \ref{fig:vorticityFields}).}
    \label{fig:LEVcontribution}
\end{figure}

The ratios quantify the signed contribution of the LEV-induced impulse to the measured force component, rather than a full decomposition of the total force. The distinction in vortex coherence is reflected in the LEV-derived to experimentally measured force ratios shown in figure \ref{fig:LEVcontribution}. Here, $F_T$ and $F_L$ denote forces in the shrimp frame of reference, and the LEV-derived forces are impulse-based estimates obtained from PIV-resolved vortex circulation.

For thrust ($F_T$), the LEV-derived contribution remains small across all cupping angles and is near zero or slightly negative for several $\zeta$ values. This indicates that the LEV plays at most a secondary role in the streamwise force balance, and in some cases contributes oppositely to thrust, producing net drag. Net thrust generation therefore primarily arises from the overall appendage motion and resistive force production during the power stroke, rather than from LEV-induced forcing alone. In contrast, the lift component ($F_L$) exhibits substantially larger variability in the LEV/Exp ratios, with the largest positive values occurring at intermediate cupping angles ($\zeta \approx 35^\circ$–$50^\circ$). In this regime, coherent LEV attachment is observed on the exopodite (figure \ref{fig:vorticityFields}), with the vortex core remaining spatially located near the leading edge and convecting with the appendage rather than detaching into the wake. The sign changes in $F_y^{LEV}/F_y^{exp}$ across $\zeta$ reflect the sensitivity of the LEV contribution to vortex timing and position relative to the appendage, rather than implying independent variation of lift and thrust over the stroke. Overall, these results support the interpretation that LEV dynamics contribute most consistently to lift enhancement at intermediate cupping angles, while their contribution to thrust is comparatively limited.

\section{Discussion and concluding remarks}

The cupping angle, $\zeta$, alters the orientation of the exopodite relative to the flow, changing the surface component normal to the incoming velocity and influencing the partitioning of force between drag-based thrust and lift. At intermediate $\zeta$ ($\zeta = 35^\circ$), rapid exopodite abduction early in the power stroke increases projected area when streamwise velocity is highest and provides sufficient time for a coherent leading-edge vortex to develop and remain attached during peak force generation. At smaller and larger cupping angles the flow either remains predominantly drag-dominated or the vortex coherence is degraded, limiting lift augmentation. These results demonstrate that the cupping angle governs the balance between drag-dominated and lift-augmented propulsion and has important ecological implications.

Shrimp do not operate at a cupping angle range that maximizes thrust or lift independently; instead, they maintain high total force generation while achieving a lift-to-thrust ratio sufficient to offset body weight. A unity $L/T$ ratio is not achieved at this angle, reinforcing that shrimp do not require equal lift and thrust to meet force balance requirements. Operating at $\zeta = 35^{\circ}$ thus reflects a functional compromise between force magnitude and kinematic effectiveness: rapid exopodite deployment promotes sustained leading-edge vortex development and lift generation, while overall force output remains near its maximum. Different swimming behaviours impose distinct force requirements. For example, forward propulsion may emphasize thrust, but hovering or inverted swimming may require greater lift. Negative lift at shallow $\zeta$ does not necessarily indicate reduced performance; at inverted orientations this force may contribute lift to the vertical force balance. Shrimp often swim upside down while grazing near the ocean surface, where such lift polarity could help resist sinking. Moreover, asynchronous activation of multiple pleopods can compensate for suboptimal performance of a single appendage, allowing force deficits at individual appendages to be compensated at the system level.

The leading-edge vortex contributes to the temporal variation of force production. When the vortex remains attached during appendage acceleration, circulation builds in phase with increasing projected area, reinforcing lift at the moment of peak hydrodynamic loading. This coupling between kinematics and vortex development suggests that intermediate cupping angles synchronize vortex growth with stroke timing. At higher cupping angles, this synchronization breaks down: either circulation is insufficient to meaningfully augment lift, or separation disrupts force generation. Hence, shrimp operate in a regime where unsteady vortex attachment is stabilized by appendage orientation and deployment rate, enabling lift augmentation without sacrificing thrust continuity. Rather than functioning as purely drag-based paddles, pleopods behave as hybrid propulsors that exploit both drag- and lift-based mechanisms to optimize swimming. An additional feature of the force evolution with time supports this interpretation. At cupping angles where a coherent LEV remains attached through the deceleration portion of the power stroke, a secondary thrust peak is observed. This feature is absent or substantially weakened at angles where vortex attachment is degraded. The timing of this second thrust maximum suggests that circulation generated during the acceleration phase is not dissipated immediately, but is instead reoriented as the appendage rotates to contribute to thrust.

The reduced-order model explains why cupping is an effective control parameter in metachronal swimming. Despite its simplicity and use of cycle-averaged coefficients, the model captures the dominant force trends across $\zeta$ and shows that force is drag-dominated, with inertial contributions playing a secondary role. By adjusting $\zeta$, shrimp modulate the relative contributions of thrust and lift without requiring substantial changes in stroke amplitude or frequency. For engineered systems, a geometric appendage configuration may offer a practical control strategy in multi-appendage propulsion, enabling shifts in force balance through structural design rather than increased actuation complexity. In this sense, cupping functions as an additional degree of freedom complementing kinematics by directly tuning force partitioning.

%In shrimp, faster adduction is likely facilitated by elastic recoil of the muscle-tendon system of the ramal joint \citep{hertzler_pleonal_2009}, which stores energy during exopodite abduction due to hydrodynamic loading in the power stroke. Similar passive–elastic coupling has been reported in other amphipods \citep{boudrias2002pleopods}, supporting the interpretation that pleopod deployment is largely fluid-driven. Kinematic data from \textit{P. paludosus} are consistent with this mechanism, as reductions in $\gamma$ occur while the posterior surface remains loaded, indicating that recoil-assisted adduction begins before passive reversal would occur from anterior loading alone. By relying on fluid–structure interaction to deploy the exopodite, shrimp can maintain effective thrust and lift generation without continuous effort, potentially reducing energetic cost while preserving maneuverability.

Our results establish the cupping angle, $\zeta$, as a key parameter governing transitions in force production, vortex formation, and stroke timing. The passive implementation in our robotic model demonstrates that effective force modulation can be achieved without complex actuation. Biologically, our results highlight the functional significance of biramous appendages for swimming. From an engineering perspective, tuning $\zeta$ adds a complementary control beyond kinematics by shifting the force balance to prioritize thrust, lift, or drag reduction at either the system or appendage level. Although $\zeta$ was held constant throughout each stroke in this study, dynamically varying cupping could further enhance maneuverability and gait transitions under unsteady flow. Future work should examine the role of dynamic $\zeta$ in free-swimming robots, particularly its potential to stabilize leading-edge vortices by optimizing the effective angle of incidence of the exopodite relative to local flow.

\backsection[Acknowledgements]{We are especially grateful to Dan Harris for his advice and many helpful discussions. We also thank Kenny Breuer, Margaret Byron, and Krista Soderlund for their valuable feedback, which strengthened the quality of this manuscript. M.M.W. and N.T. were supported by the ONR Bio-inspired Autonomous Systems program (N000142412662).}

\backsection[Declaration of interests]{The authors report no conflict of interest.}

%Use of the above commands will create a bibliography using the .bib file. Shown below is a bibliography built from individual items.

\bibliographystyle{jfm}
\bibliography{jfm}

\end{document}

\maketitle

\section*{Supplementary Figures}

%\begin{figure}
 %   \centering
  %  \includegraphics[width=0.5\linewidth]{Supplementary Figures/force_transducer.png}
  %  \caption{\textbf{Force frame of reference}. The forces are shown in three frames of reference: laboratory, shrimp, and global frames. Forces on the rotating paddle (blue) are measured in two directions, one that is perpendicular ($F_{\perp}$ or $F_y$) and one that is parallel to the paddle ($F_x$ or $F_{//}$). In the shrimp frame of reference, $F_n$ is the force normal to the swimming direction, and $F_t$ is the force tangential to the swimming direction. In the global frame of reference, lift is the force pointing vertically, and thrust is the force pointing horizontally. $\psi$ is the angle between the body axis and the endopodite. The paddle illustration shows the top view in the laboratory frame of reference, with the axis of rotation and endopodite labeled. A shrimp lateral view shows the paddle orientation in the shrimp frame of reference.}
   % \label{fig:frameOFreference}
%\end{figure}

\begin{figure}
    \centering
    \includegraphics[width=1\linewidth]{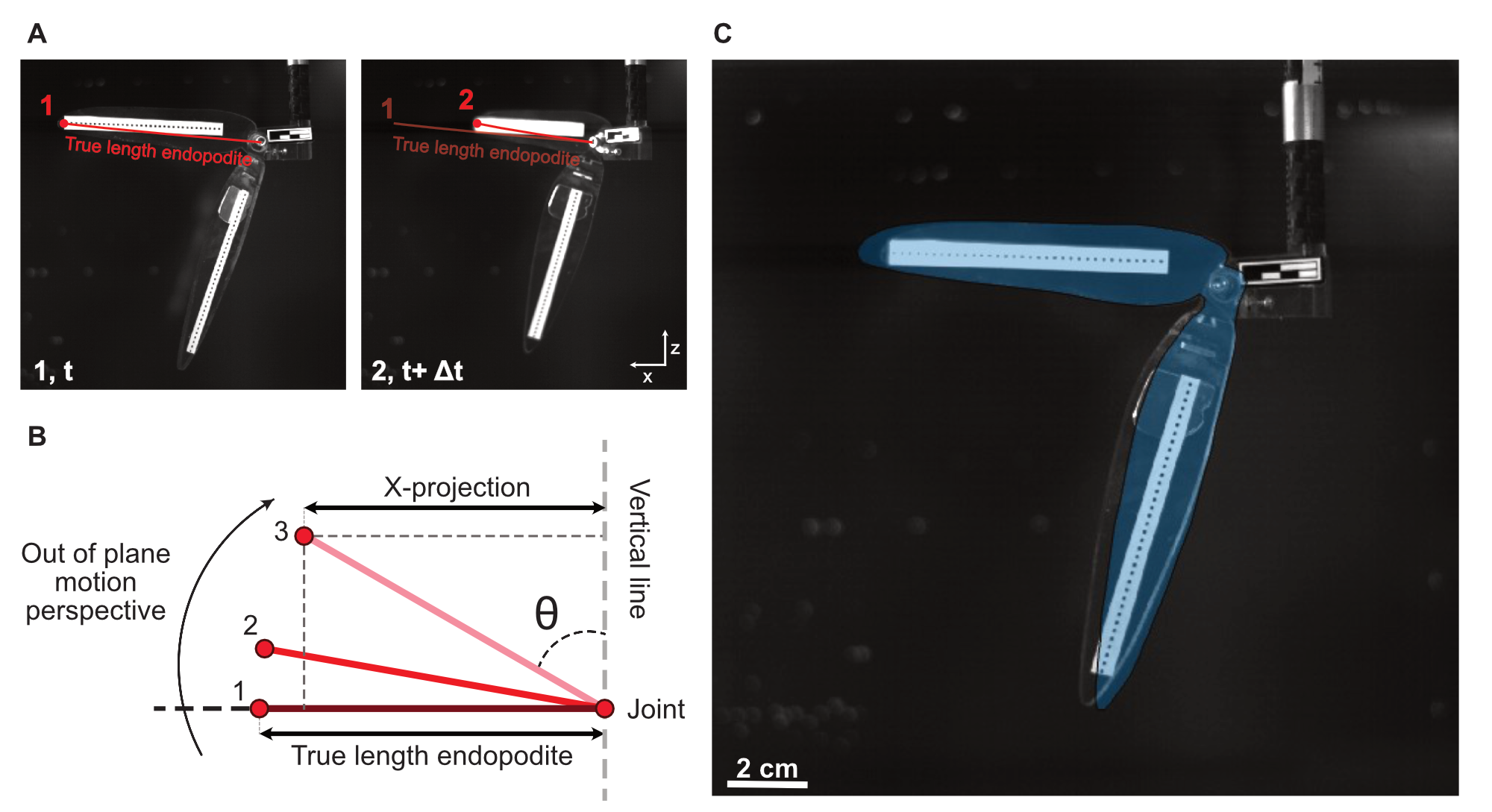}
    \caption{\textbf{Tracking locations on the model used during kinematics experiments.} The angle between the endopodite and exopodite,$\gamma (t)$, is the angle between lines passing through points 1,2 and 3,2, respectively. Tracked points correspond to landmarks on the 3D model, where dotted tape was used to track relevant points for kinematics analysis. The model is outlined to increase contrast with the background. In the frame of reference of the model, $L$ is the actual exopodite length, measured from Points 2 to 3. The points closest and furthest from the camera are also shown to illustrate the difference in angle with changing perspective. These parameters determine the instantaneous $\gamma$ while considering the effect resulting from the fixed perspective.}
    \label{fig:changeOFarea}
\end{figure}

\begin{figure}
    \centering
     \includegraphics[scale=0.9]{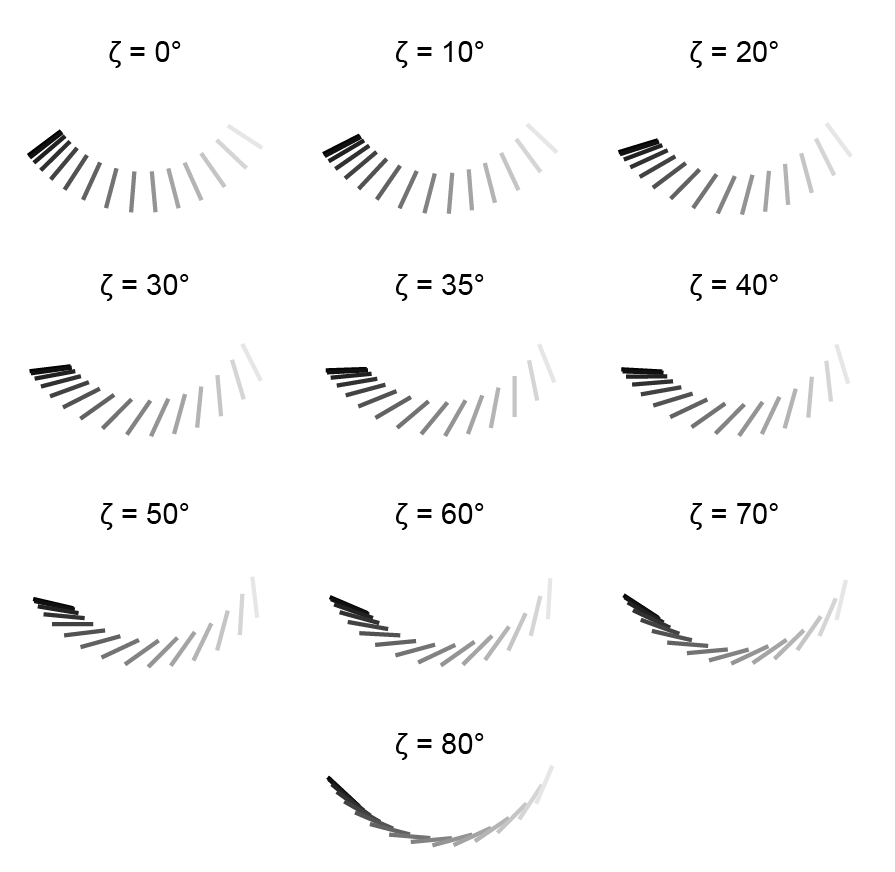}
    \caption{\textbf{Effective angle of attack ($\mathbf{\delta}$) over the power stroke.} The position of the exopodite throughout the power stroke is shown here, with the bars representing a cross-section of the exopodite in a lateral view. The kinematics are governed by the angle $\psi$, the angle between the body axis and the plane that bisects the endopodite. The cupping angle, $\zeta$, is applied to the exopodite, and the overall angle of the exopodite is $\delta = \psi - \zeta$. The effective AoA for all values of $\zeta$ is shown here in a side view bisecting the extended exopodite. The beginning of the power stroke is represented by the black bars, and the lighter colors signify the time progression through the power stroke. The lightest grey color signifies the end of the power stroke. Uneven spacing between profiles represents areas of acceleration and deceleration.}
    \label{fig:compound_angle_for_all_zeta}
\end{figure}

\begin{figure}
    \centering
    \includegraphics[width=0.8\linewidth]{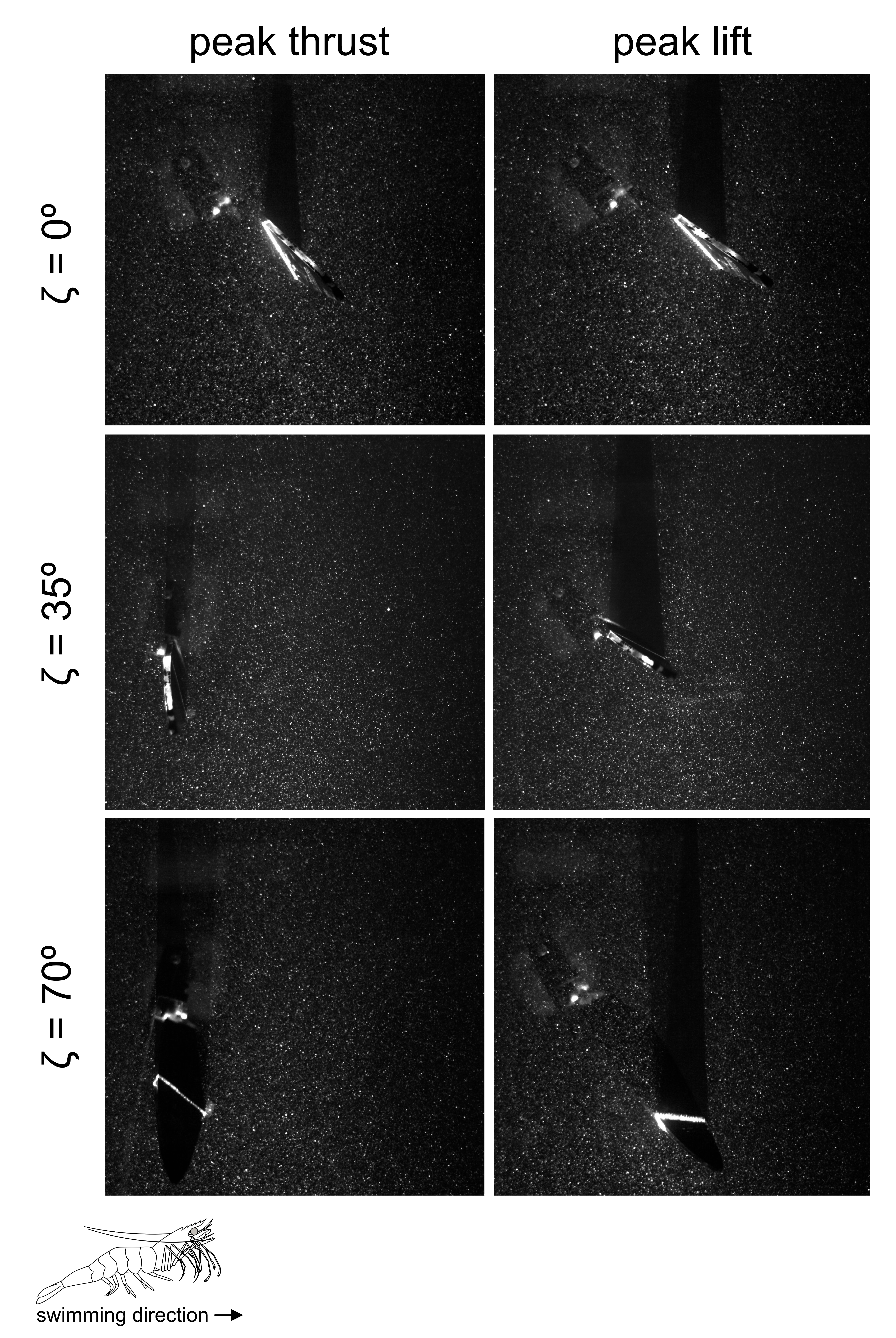}
    \caption{Raw particle image velocimetry (PIV) images corresponding to the vorticity and streamline fields shown in Figure 8, at maximum measured thrust (left column) and maximum measured lift (right column) for three cupping angles: $\zeta = 0^\circ$, $35^\circ$, and $70^\circ$ (top to bottom). These images show the particle distribution within the laser sheet intersecting the exopodite cross-section, prior to velocity and vorticity computation. The exopodite outline is not overlaid here to preserve the raw image.}
    \label{fig:vortRAW}
\end{figure}